\documentclass[usenatbib]{mn2e}

\usepackage{amssymb,amsmath}
\usepackage{graphicx}
\usepackage{ctable}

\usepackage{abbreviations}

\newcommand{\be}{\begin{equation}}
\newcommand{\bb}{\begin{equation}}
\newcommand{\ee}{\end{equation}}
\newcommand{\s}[1]{_{\rm{#1}}}
\def\la{\buildrel < \over {_{\sim}}}
\def\ga{\buildrel > \over {_{\sim}}}

\newcommand\eg{\textit{e.g.,\ }}

\newcommand{\gp}{\gamma_{\rm p}}

\graphicspath{ {./}{Figures/} }

\title[Diffusion in PWNe]{Diffusion in Pulsar Wind Nebulae: An Investigation using Magnetohydrodynamic and Particle Transport Models}
\author[O. Porth, M.~J. Vorster, M. Lyutikov  and N.~E. Engelbrecht]{
	O. Porth,$^{1,2}$\thanks{E-mail: porth@th.physik.uni-frankfurt.de}	
	M.~J. Vorster$^{3,4}$\thanks{E-mail: michael.j.vorster@gmail.com} 
	M. Lyutikov,$^{3}$\thanks{E-mail: lyutikov@purdue.edu} and
	N.~E. Engelbrecht,$^{4}$\thanks{E-mail: 12580996@nwu.ac.za}
\\
	{$^1$ \it Institute for Theoretical Physics, Frankfurt am Main, D-60438, Germany} 
	\\
	{$^2$ \it Department of Applied Mathematics, The University of Leeds, Leeds LS2 9JT, UK}
\\
	{$^3$ \it Department of Physics, Purdue University, 525 Northwestern Avenue, West Lafayette, IN 47907-2036, USA}
\\
	{$^4$ \it Centre for Space Research, North-West University, Potchefstroom, 2520, South Africa}
}

\begin{document}

\date{Accepted -----. Received -----}

\maketitle

\label{firstpage}

\begin{abstract}
We study the transport of high-energy particles in pulsar wind nebulae (PWN) using three-dimensional MHD (see \cite{Porth2014} for details)  and test-particle simulations, as well as a Fokker-Planck particle transport model.  The latter includes radiative and adiabatic losses, diffusion, and advection on the background flow of the simulated MHD nebula.   By combining the models, the spatial evolution of flux and photon index of the X-ray synchrotron emission is modelled for the three nebulae G21.5-0.9, the inner regions of Vela, and 3C 58, thereby allowing us to derive governing parameters: the magnetic field strength, average flow velocity and spatial diffusion coefficient.  For comparison, the nebulae are also modelled with the semi-analytic \cite{Kennel1984a} model but the \cite{Porth2014} model generally yields better fits to the observational data.

We find that high velocity fluctuations in the turbulent nebula (downstream of the termination shock) give rise to efficient diffusive transport of particles, with average P\'eclet number close to unity, indicating that both advection and diffusion play an important role in particle transport.  We find that the diffusive transport coefficient  of the order of $\sim2\times 10^{27} (L_{\rm s}/0.42\rm Ly) cm^{2}s^{-1}$  ($L_{\rm s}$ is the size of the termination shock) is independent of energy up to extreme particle Lorentz factors of $\gamma_{p}\sim10^{10}$.  
\end{abstract}


\section{Introduction} \label{sec:introduction}
 
Pulsars produce highly relativistic winds that consist of electrons and positrons,  and frozen-in  magnetic field \citep[e.g.,][]{Kirk2009}. 
Where the ram pressure of this magnetized wind equals the confining pressure of the ambient medium, a termination shock is formed \citep{Rees1974,Kennel1984a}. Charged particles are accelerated at the termination shock  \citep[e.g.,][]{Kennel1984b,Reynolds1984}, presumably via  the Fermi acceleration mechanism but also magnetic reconnection of the striped pulsar wind could play an important role \citep[see, e.g.,][]{Sironi2011}.     
Downstream of the termination shock, relativistic leptons produce synchrotron radiation from radio to X-ray wavelengths.  Particles  are continuously transported away from the termination shock by the downstream flow, inflating a luminous nebula around the pulsar commonly known as a pulsar wind nebula (PWN).  

Overall,  the synchrotron energy spectra can be described using a combination of  power laws \citep[see, e.g.,][]{Dejager2009}.  Spatially-resolved X-ray observations of PWNe show that the photon index $\Gamma$ evolves as a function of distance from the pulsar  \citep[see, e.g.,][]{Slane2000, Bocchino2001, Mangano2005, Schock2010}.  
As shown by e.g. \cite{Vorster2013a}, the evolution of the lepton spectrum is not only determined by the energy loss rate, but also by the rate at which particles are transported towards the outer parts of the nebula.  While particle evolution models of PWNe typically only take convection into account, it has been suggested that diffusion may also play an important role in the transport process \citep[see, e.g.,][]{Gaensler2006}.  This idea is supported by the results of \cite{Tang2012}, who found that they were able to better model the evolution of $\Gamma$ observed for G21.5-0.9 and 3C 58 by including both convection and diffusion in their model, in contrast to only including convection.  

The exact nature of diffusion is still an open question, with both kinetic and turbulent processes at its core.  Adopting a kinetic view, one expects diffusion to be related to the magnetic field geometry, while there exists some uncertainty as to the correct topology to use for PWNe.  If one assumes the purely toroidal magnetic field from the classic \cite{Kennel1984a} model, then all outward diffusion is perpendicular to the field lines which renders kinetic diffusive transport inefficient.  
In contrast to this simple picture, observations of filamentary structures in the outer regions of Crab Nebula \cite[e.g.][and references therein]{hester2008} support the presence of a poloidal field component that wraps around and along the filaments.  
This impacts on the transport of particles, which can then slide outwardly along the field lines.  In fact, deep Chandra X-ray observations by \cite{SewardSewardTucker2006} find indications that spectral index variations along filaments are less pronounced than in the cross-field direction, highlighting the importance of the magnetic topology in particle transport.  

There is now little doubt that the prominent outer filaments of Crab nebula are due to the Rayleigh-Taylor instability (RTI) of the interface between the PWN and the supernova remnant \citep{ChevalierChevalierGull1975,jun1998,porthRT2014, BietenholzNugent2015}.  
In general, the RTI can be suppressed for strong and ordered field \citep[e.g.][]{bucciantini2004} or when the interface has stopped accelerating in old PWNe with age of $\sim10^4\rm yr$ \citep[see also][]{ChevalierReynolds2011}.  Thus is is uncertain whether the RTI on its own can account for the destruction of ordered field and seed turbulence throughout the nebula.  

However, fluid instabilities in the bulk of the nebula, triggered by the causally connected PWN jet and equatorial shear-flow contribute to a destruction of the ordered toroidal field that is injected with the relativistic wind \citep[e.g.][]{begelman1998,Mizuno:2011aa,Porth2014}.  
Perhaps more than by changing the geometry, fluid instabilities can directly enhance particle transport through driving a turbulent cascade in which particles are transported diffusively as we will show in \S \ref{sec:diffusion}.  
  
The present paper aims to address two specific points: the first is to use the MHD results of \cite{Porth2014} to study particle transport and derive diffusion coefficients for PWNe.  The second is to further investigate the role of diffusion in the evolution of the leptons spectra by extending the results of \cite{Tang2012}.  These authors focused on modelling the spatial evolution of $\Gamma$, while neglecting the corresponding evolution of the X-ray synchrotron flux.  Naively one might expect that a good fit to $\Gamma$ would automatically produce a good fit to the flux.  However, \cite{Vorster2013t} found that the evolution of $\Gamma$ is, to a degree, determined by the strength of the magnetic field, the convection velocity, and the diffusion coefficient relative to each other.  It is therefore possible to find a similar evolution of $\Gamma$ by suitably varying the above-mentioned parameters.  The paper therefore addresses this issue by taking into account the spatial evolution of the flux in the modelling, where possible.  
    
While the magnetic field and flow velocity calculated by \citet[][hereafter PKK14]{Porth2014} are the primary choice for investigating the role of diffusion, it is useful to compare the results with those found from using the classic model developed by \citet[][hereafter KC84]{Kennel1984a}, where the spatial dependence of the magnetic field and flow velocity are determined by the ratio $\sigma$ of magnetic to particle energy in the PWN.  The data fitting will therefore allow one to derive the value of this important parameter within the context of the KC84 model.  Thus far this value has only been rigorously derived for the Crab Nebula \citep{Kennel1984b}.  

The rest of the paper is structured as follows:  Sec. \ref{sec:diffusion} presents the diffusion coefficients derived from the PKK14 simulations. Section \ref{sec:model} introduces the particle transport model that is used to model the X-ray data, with the modelling results presented in Sec. \ref{sec:results}.  Lastly, a discussion of the results, together with the main conclusion drawn from the modelling can be found in Sec. \ref{sec:conclusions}.

\section{Diffusion in Pulsar Wind Nebulae}\label{sec:diffusion}

The coefficients of ordinary diffusion can be defined as
\begin{align}\label{eq:Dxy}
D_{xy}(\Delta t) = \frac{\langle \Delta x \Delta y\rangle}{2 \Delta t},  
\end{align}
where the average is taken over the differences in positions $\Delta x= x(t+\Delta t)-x(t)$.  For the diffusion approximation to hold, many small angle deflections within $\Delta t$ are required, and result in a stochastic behavior of the particles.  It is therefore required that $\Delta t\gg \tau$, where $\tau$ is the collision time scale.  Under these circumstances $D_{xy}$ becomes independent of $\Delta t$.  On the other hand, for anomalous diffusion the mean squared displacement follows a more general relation $\langle \Delta x^2\rangle\propto \Delta t^{\beta}$, with $\beta\ne1$ such that the running diffusion coefficient, as given by Eq. (\ref{eq:Dxy}), will show an inherent time-dependence.  Whether particle transport in simulated PWN obeys relation (\ref{eq:Dxy}) or a more general law will be investigated in Sec. \ref{sec:numerical}.  We first discuss two diffusion processes likely to occur in PWN and then move to the numerical results.  

\subsection{Ordered field: Bohm diffusion}
\label{Bohm}

Analytical \citep{Kennel1984a}, as well as 2-dimensional MHD models of PWNe adopt a purely toroidal field geometry.  Radial transport  can then  occur either  due to (turbulent) advection or kinetic cross-field diffusion.  A useful, upper limit for cross-field diffusion is provided by the empirical Bohm-law 
\begin{align}
D^{\rm B} = \frac{1}{3}r_g^2 \omega_g = 1.7 \times 10^{26} \left(\frac{\gp}{10^9}\right) \left(\frac{B}{100{\rm \mu G}}\right)^{-1}\ \rm cm^2~s^{-1} \label{eq:bohm}
\end{align}
with diffusion time
\begin{align}
t_{\rm B} = \frac{L^2}{2D^{\rm B}} = 3 \times 10^{3} \left(\frac{\gp}{10^9}\right)^{-1} \left(\frac{B}{100{\rm \mu G}}\right) \left(\frac{L}{2\rm pc}\right)^2 \rm yrs
\end{align}
Even for PeV particles this is larger than the age of the Crab Nebula. 

On the other hand, the advective time scale $t_a$ can be estimated from the flow velocity in the nebula.  In the laminar model of \citep{Kennel1984a}, the flow velocity monotonously connects the expansion speed $v_n$ with the fast flow near the termination shock $v_f\sim c/\sqrt{3}$, hence
\begin{align}
4\times 10^{8} \left(\frac{L}{2\rm pc}\right) \left(\frac{v_f}{0.5 c}\right)^{-1}\ \rm s\le t_a \\
 \le 4\times10^{10} \left(\frac{L}{2\rm pc}\right) \left(\frac{v_n}{1500 \rm km~s^{-1}}\right)^{-1}\ \rm s
\end{align}
which is shorter than $t_{\rm B}$ for particles up to 1 PeV. Thus, kinetic particle diffusion with $D^{\rm B}$  is negligible in PWNe.

\subsection{Diffusion in a turbulent flow}\label{sec:eddydiff}

Next to the field geometry, an important factor in particle diffusion in PWNe is the presence of vigorous turbulence. In this case even particles tied to a given field line experience spacial diffusion due to the effect of field line wandering \citep[\eg][]{SaluSaluMontgomery1977,1995PhRvL..75.2136M,1994plas.conf.....K}.  
This ``frozen-in'' diffusion should not be mistaken with energy dependent kinetic diffusion due to wave-particle interactions as it is often encountered in theories for cosmic ray diffusion \citep[e.g.][]{Shalchi2009}.  

Let us now provide a back of the envelope estimate for the effective diffusivity that is connected to the turbulent nebula flow.  
The role of the velocity field can be easily seen at the simplified axisymmetric system where $\mathbf{B}=B \mathbf{\hat{e}}_{\phi},\ \mathbf{v}=v_{r}\mathbf{\hat{e}_{r}}+v_{\theta}\mathbf{\hat{e}_{\theta}}$.  This reflects our plasma injection conditions in good approximation to the pulsar wind far away from the source.  
Here, the drift velocity $\mathbf{v}_{D}=\mathbf{E\times B}/B^{2}$ reduces to the fluid velocity $\mathbf{v}_{D}=\mathbf{v}$ and to first order, particles are simply advected with the flow. 

For the largest eddies on the scale of the termination shock $L_{\rm s}$ (the driving scale of the turbulence), we obtain a typical velocity of $v_f\sim0.5 c$ close to the sound-speed in an ultra-relativistic gas $c_{\rm s}=c/\sqrt{3}$.  The corresponding eddy turnover time is
\begin{align}
t_{\rm Ls} = \frac{L_{\rm s}}{v_f} = 2.8\times 10^7 \left(\frac{L_{\rm s}}{0.42\rm Ly}\right)\left(\frac{v_f}{0.5c}\right)^{-1}\rm s
\end{align}
which is much shorter than the nebula age.  Thus interpreting $t_{\rm Ls}$ as the collision time, we have $\Delta t \gg \tau$ and expect a diffusive transport of particles tied to the eddies.  If the nebula expands self-similarly, the latter will remain true during all of its evolution.  
The effective eddy diffusion coefficient on the scale $L_{\rm s}$ is hence estimated with 
\begin{align}
D^{\rm E}_{\rm Ls} = \frac{1}{3} v_f L_{\rm s} = 2.1 \times 10^{27}\left(\frac{v_f}{0.5c}\right)\left(\frac{L_{\rm s}}{0.42\rm Ly}\right)\ \rm cm^2~s^{-1}\, .\label{eq:DE}
\end{align}
We again estimate the diffusion timescale for typical parameters of a young PWN, 
\begin{align}
t_{\rm E} = \frac{L^2}{2D^{\rm E}} = 8.57\times 10^{9} \left(\frac{v_f}{0.5c}\right)^{-1}\left(\frac{L_{\rm s}}{0.42 \rm Ly}\right)^{-1}\left(\frac{L}{2\rm pc}\right)^2\ \rm s
\end{align}
which is less than 300 years.  This corresponds to the ``escape time'' of particles injected at the termination shock.  

Adopting a \cite{Kolmogorov1941a} scaling for the velocities on the two scales $\lambda L$ and $L$: $v_{\lambda L}\sim \lambda^{1/3} v_{L}$, we obtain the scalings for the eddy turnover time $t_{\lambda L}\sim \lambda^{2/3} t_{L}$ and for the diffusion coefficient $D^E_{\lambda L}\sim \lambda^{4/3} D^E_{L}$.  One can see that the largest scales dominate the diffusion process which is why we could neglect the contribution of smaller scales in our estimate (\ref{eq:DE}).  
Indeed, if we define a scale-averaged diffusion coefficient 
\begin{align}
\langle D^E \rangle_{\lambda0} \equiv \frac{1}{1-\lambda_0}\int_{\lambda0}^1 \lambda^{4/3} D_{\rm Ls}^E d\lambda \label{eq:Dlambda}
\end{align}
we see that the effect of allowing smaller scales is rather small:  in the limit $\lambda_0\to 0$, we merely obtain $\langle D^E \rangle_{\lambda}\simeq 0.43 D_{Ls}^E$.  

Note that the diffusion time in terms of the eddy turnover time is just $t_{\rm E}/t_{\rm Ls}=3/2 (L/L_s)^2$ which is roughly conserved in the self-similar expansion of the nebula.  In Crab, we have $L/L_s\sim 14$, thus particles witness several hundred turnover times before escaping from the system.   

The coefficient given by Eq. (\ref{eq:DE}) is quite universal as its governing dependence is just the termination shock size -- it could thus easily be adopted to other PWNe.  
Since $D^{\rm E}>D^{\rm B}$, for energies up to $\sim 6\rm PeV$, we do not expect an energy dependence of the diffusion coefficient for the large majority of particles.  In particular, no energy dependence for Xray-synchrotron emitting particles is expected.

\subsection{Test-particle diffusion in 3D simulations of PWN}\label{sec:numerical}

To estimate the diffusion coefficient in the MHD simulations of \cite{Porth2014}, we follow orbits of test-particles in the MHD flow, similar to the approach by \cite{WisniewskiSpanier2012} in application to heliospheric plasma.  Mono-energetic electrons are injected uniformly and isotropically into a simulation snapshot and the particles are advanced according to the Lorentz-force resulting from the MHD fields $\mathbf{E,B}$ \citep[e.g.][]{landau1960}:
\begin{align}
\frac{d \mathbf{u}}{dt} = \frac{q}{mc} \left( \mathbf{E}+\frac{\mathbf{u\times B}}{c\gamma_{\rm p}}\right) + \mathbf{g} \label{eq:Lorentz}
\end{align}
where $\mathbf{u}=\gamma_{\rm p}\mathbf{v}/c$ is the particles four-velocity and $q/m$ signifies the electron charge to mass-ratio.  Here we omit the radiation reaction force and set $\mathbf{g=0}$.  Eq. (\ref{eq:Lorentz}) is advanced with a second order symplectic Boris scheme \citep[e.g.][]{BirdsallLangdon1991} where the fields $\mathbf{E,B}$ at the particles position are obtained via linear interpolations in space and time between the fluid steps.  For numerical stability, each particle is advanced with individual time step ensuring that one gyration is resolved by at least 60 steps.  
This particle treatment is implemented in MPI-AMRVAC\footnote{https://gitlab.com/mpi-amrvac/amrvac} \citep{Keppens2012718,PorthXia2014} allowing for massively parallel calculations and adaptive mesh refinement.  To obtain sufficient statistics, we follow the orbits of $5\times 10^5$ particles.  
The resulting diffusion coefficients can then be measured for different time intervals $\Delta t$ according to Eq. (\ref{eq:Dxy}).  Here we restrict ourselves to the models with parameters as ``B3D'' of PKK14.  Thus we adopt an obliqueness angle of $\alpha=45^\circ$, and an initial wind-magnetisation $\sigma_0=1$.  

\begin{figure}
\begin{center}
\includegraphics[height=6cm]{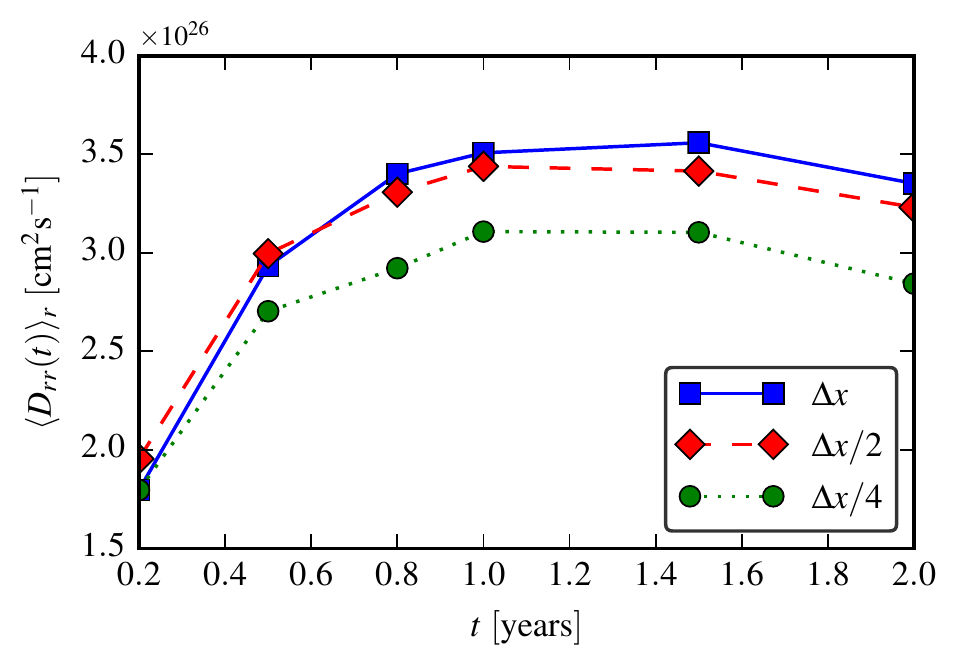}
\includegraphics[height=6cm]{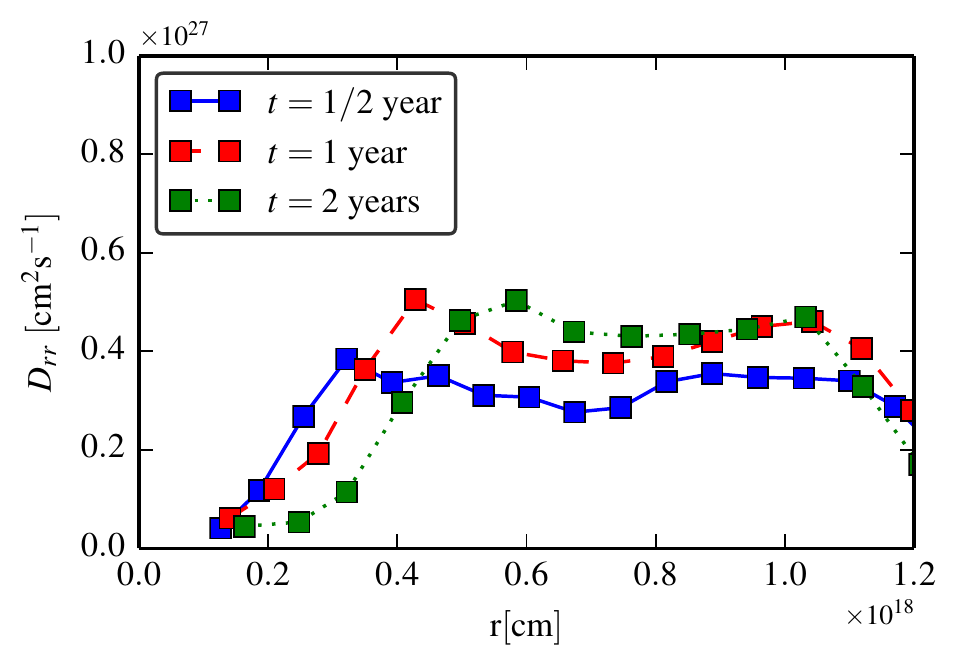}
\caption{\textit{Top:} Convergence in time of the mean radial diffusion coefficient $\langle D_{rr}(t)\rangle$ for particles with $\gamma_{\rm p}=10^8$ for increasing resolutions of the MHD flow, where $\Delta x=2\times 10^{16}\rm cm$.  Convergence in time is obtained after $\approx1$ year.  After 1.5 years, the estimated diffusion coefficient decreases again, most likely due to particle escape. 
\textit{Bottom:} Radial profiles of the radial diffusion coefficient for $\gamma_{\rm p}=10^8$ for various time intervals $t$}
\label{fig:drrvstime}
\end{center}
\end{figure}

To check if normal diffusion is an accurate description for the transport process, we first investigate the convergence of the running diffusion coefficient.  
Fig. \ref{fig:drrvstime} (left panel) thus shows the mean radial diffusion coefficient obtained within the PWN volume as a function of the time-interval.  
Here $D_{rr}(\Delta t)$ is calculated by simply replacing $\Delta x \Delta y$ in Eq. (\ref{eq:Dxy}) with $\Delta r^2$ for the radial transport.  
For the relatively short averaging times employed, transport due to mean-flow advection (see also Sec. \ref{sec:PKK14}) is negligible at least in the outer parts of the nebula and has not been subtracted in the determination of $D_{rr}$.  

After an initial ballistic regime where $D_{rr}\propto t,$, convergence is obtained at approximately one year (roughly three $t_{\rm Ls}$) at a value of $\langle D_{rr}\rangle\approx 3.5\times 10^{26}\rm \ cm^2~s^{-1}$.  Taking into account the termination shock size of $L_s\simeq 0.2\rm Ly$ in the simulation, this agrees reasonably well with the expectation given by Eq. (\ref{eq:DE}).  
As the numerical resolution of the MHD simulation is increased by a factor of four, the temporally converged diffusion coefficient decreases by $14\%$ from a value of $3.5\times10^{26}\ \rm cm^2\, s^{-1}$ to $3\times10^{26}\ \rm cm^2\, s^{-1}$.  
We suspect that this moderate dependency reflects the fact that our simulations are not yet able to reach perfect convergence of the velocity power-spectrum.  

The bottom panel of Fig. \ref{fig:drrvstime} illustrates the radial profile of $D_{rr}$.  In the inner (laminar) region of the PWN, particle diffusion is suppressed with coefficient rising to a plateau reached at $\approx1/3$ of the nebula radius.  
The systematic radial advection present in the inner nebula leads to an outward shifting offset of the profiles as longer time-intervals are considered.  
At the outer edge, particle escape leads to a systematic decrease of the measured coefficient as the time-interval increases.  This cautions us to restrict the averaging times to $\approx 1$ year maximum.  

\begin{figure*}
\begin{center}
\includegraphics[height=6.5cm]{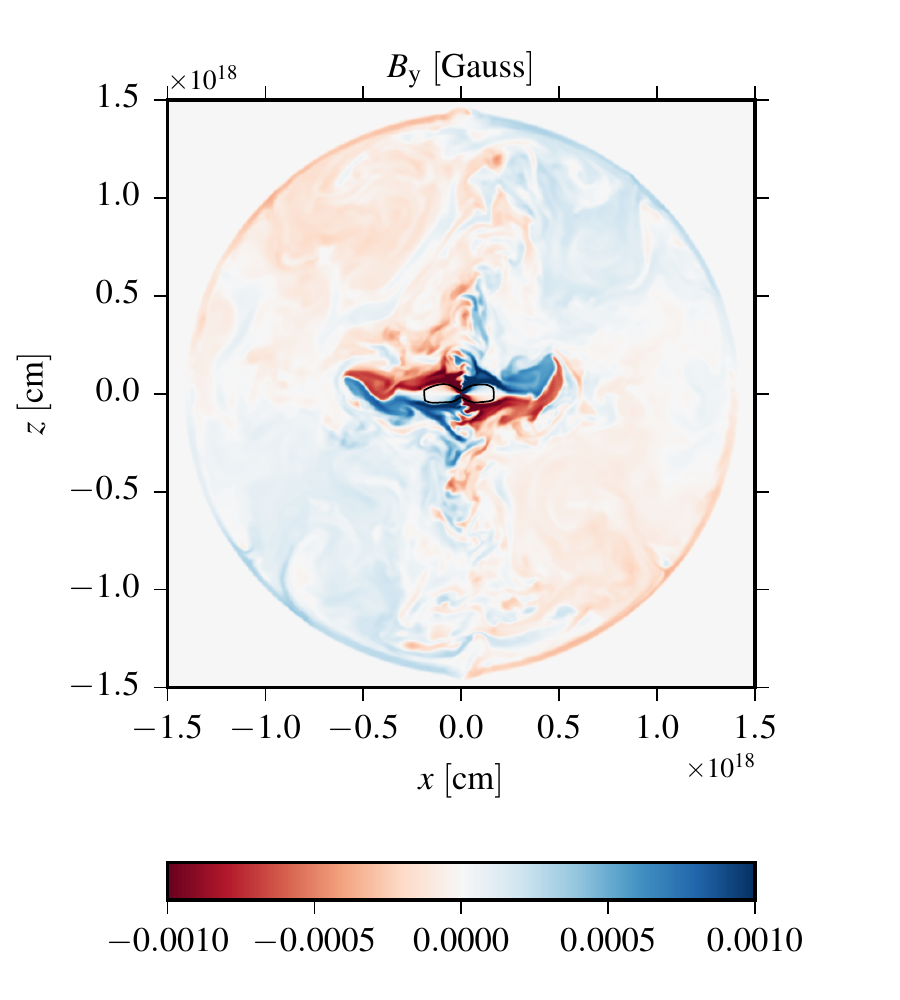}
\includegraphics[height=6.5cm]{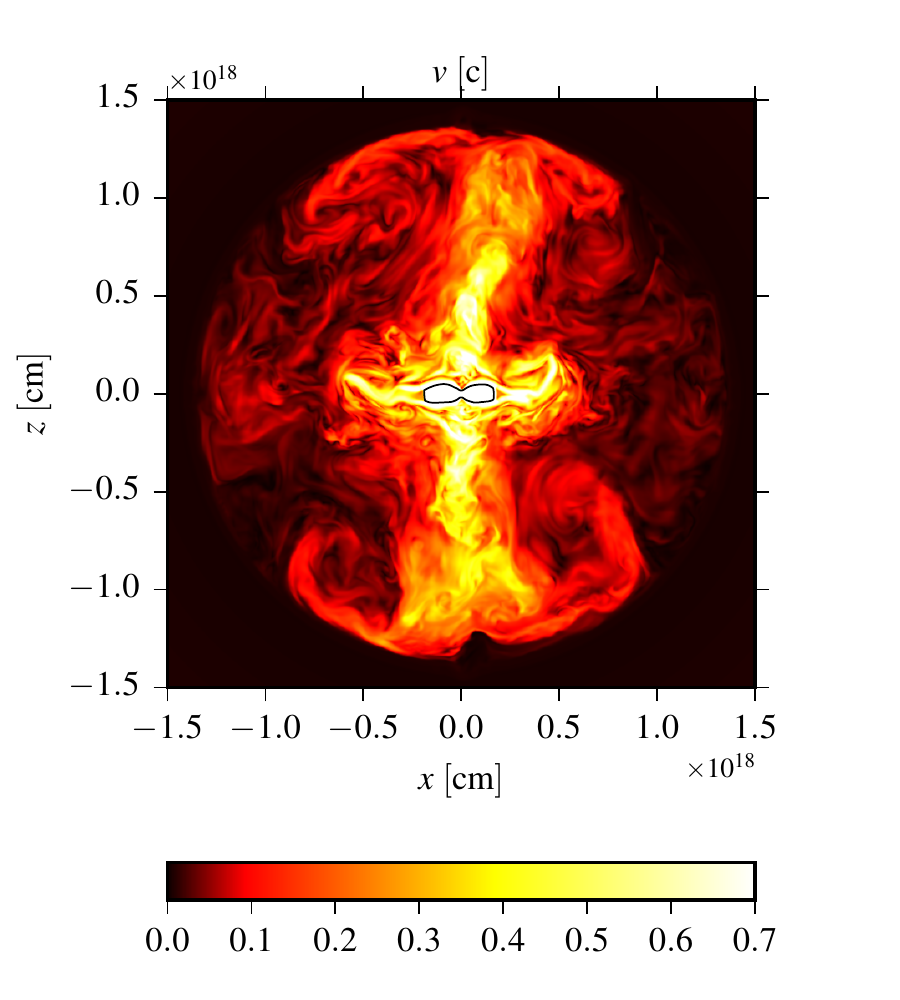}
\includegraphics[height=6.5cm]{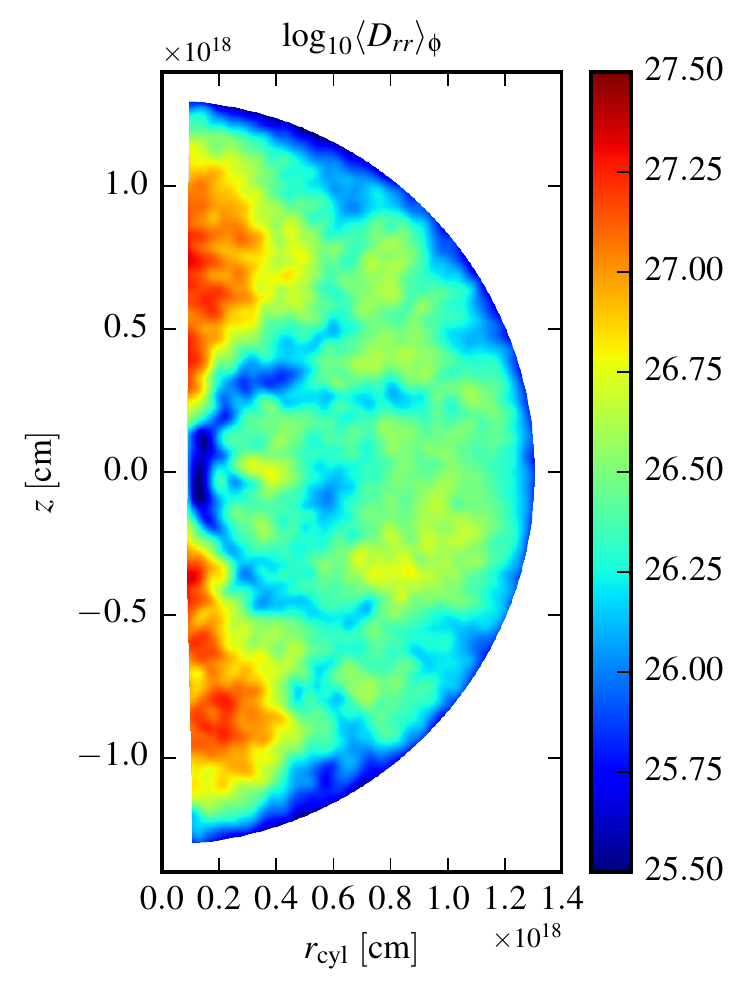}
\caption{Flow characteristics in highest resolution run with $\Delta x=5\times 10^{15}\ \rm cm$, shown in the $y=0$ plane.  
\text{Left:} Out-of-plane component of the magnetic field illustrating the formation of an equatorial current-sheet in the torus region of the nebula.  The net-polarity in the two hemispheres is still given by the injected field, indicating a toroidal guide-field.  
\textit{Middle:} Velocity magnitude illustrating development and decay of small-scale turbulent flow within 1-2 termination shock radii.  
\textit{Right:} Map of the radial diffusion coefficient as obtained from the test-particles and averaged over azimuthal direction.  High transport coefficients are realised especially in the polar plume.  The drop at the outer radius is a systematic introduced by particle escape.  }
\label{fig:Drr-rz}
\end{center}
\end{figure*}

Finally, we show flow characteristics and a map of the corresponding $\langle D_{rr}\rangle_{\phi}$ in Fig. \ref{fig:Drr-rz}.  The out-of-plane magnetic field indicates that the slowly varying striped wind (with obliqueness angle $\alpha=45^\circ$) forms a current-sheet in the torus region of the nebula.  Magnetic polarity mixes in the bulk of the nebula as can be seen at the example of the blue strand of positive flux in the north-western quadrant.  None the less, the net-polarity in each hemisphere is given by the injected one.  
Equatorial shear flow and the violently unstable polar jet generate turbulence and small scale flow within one to two termination shock radii.  This way, the instantaneous flow velocity differs substantially from the radial expansion of the average flow.  Only the latter has to match with the slow expansion velocity of the nebula.  This is a major difference to the laminar model of KC84.  

Particle transport is highest in the region of the polar plume and near the equator.  This is expected, as in these regions, violent mixing is triggered via kink-instability and fast shear-flow.  We obtain a variation of the transport coefficients by almost two orders of magnitude in the nebula where the minimum is found at intermediated polar angles of $45^{\circ}$.  

\begin{figure}
\begin{center}
\includegraphics[height=6cm]{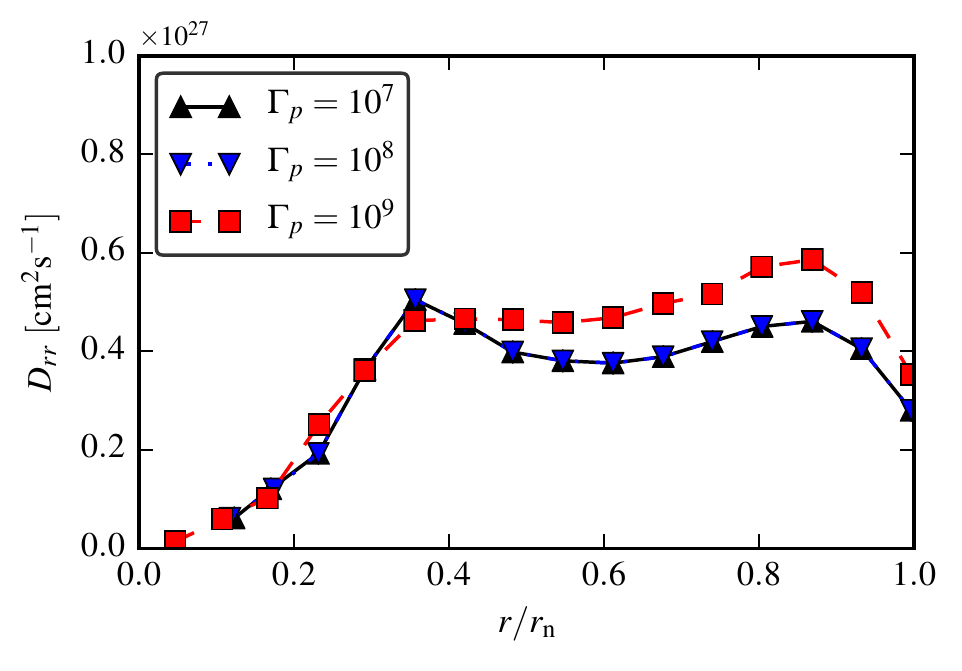}
\caption{Transport coefficient for various particle energies.  Diffusion remains energy independent until the gyroradius becomes larger than the scale $L_{\rm S}$ at $\gp >10^{9}$.  Results for $\gp=10^7$ and $\gp=10^8$ overlap almost identically.  }
\label{fig:dvsenergy}
\end{center}
\end{figure}

The variation of the local diffusion coefficients with particle energy are shown in Fig. \ref{fig:dvsenergy}.  
As expected, the spatial diffusion coefficient increases once the gyroradius
\begin{align}
  r_g=\frac{p_\perp c}{e B} = 1.7\times 10^{16} \left(\frac{\gp}{10^9}\right)\left(\frac{B}{100\mu G}\right)^{-1}\ \rm cm
\end{align}
becomes comparable to the scale $L_{\rm S}$.  At a typical magnetic field strength of $100\mu\rm G$, this is the case for ultra-relativistic particles with $\gp\approx 10^{10}$.  These unrealistically high energy particles surpass even the Crab-flare particles of $\gp\approx 1\times 10^9- 3\times 10^9$ \citep{tavani2011, abdo2011} which is why we will pay no further mind to them. 
However, for X-ray emitting particles with energies of $\gp\lesssim10^7$, we do not observe an energy dependence of the diffusion coefficient.  
It is clear that the numerical resolution of $\Delta x=2\times 10^{16}\rm cm$ does not suffice to resolve variations in the flow on the order of the gyroradius for $\gp\lesssim10^9$ and the particle transport reduces to the motion of the guiding centres.  In this sense, the lacking energy dependence comes as no surprise.  However, as the relevant transport process for X-ray emitting particles is turbulent eddy diffusion which in turn is dominated by the well resolved largest scale of the cascade (Sec. \ref{sec:eddydiff}), we expect the results obtained for $D_{rr}$ to be in the physically relevant regime.

\subsection{Characterisation of the turbulent flow}\label{sec:turbulence}

To derive quantities of interest for the turbulent transport of particles, we recover an averaged and a fluctuating part of the MHD vector-fields
\begin{eqnarray}
\langle B\rangle &=& \sqrt{\langle B_\phi\rangle^2+\langle B_r\rangle^2+\langle B_\theta\rangle^2} \\
\delta B &=& \sqrt{\delta B_r^2+\delta B_\theta^2+\delta B_\phi^2} \\
\delta B_i &=& \sqrt{\langle B_i^2\rangle - \langle B_i\rangle^2}
\end{eqnarray}
where the averages are performed over several correlation timescales and the azimuthal direction.  
Since the initial- and boundary conditions are independent of the $\phi$-coordinate, we can interpret the $\phi$-average as an ensemble average which improves the statistics of the data.  

\begin{figure*}
\begin{center}
\includegraphics[height=7.5cm]{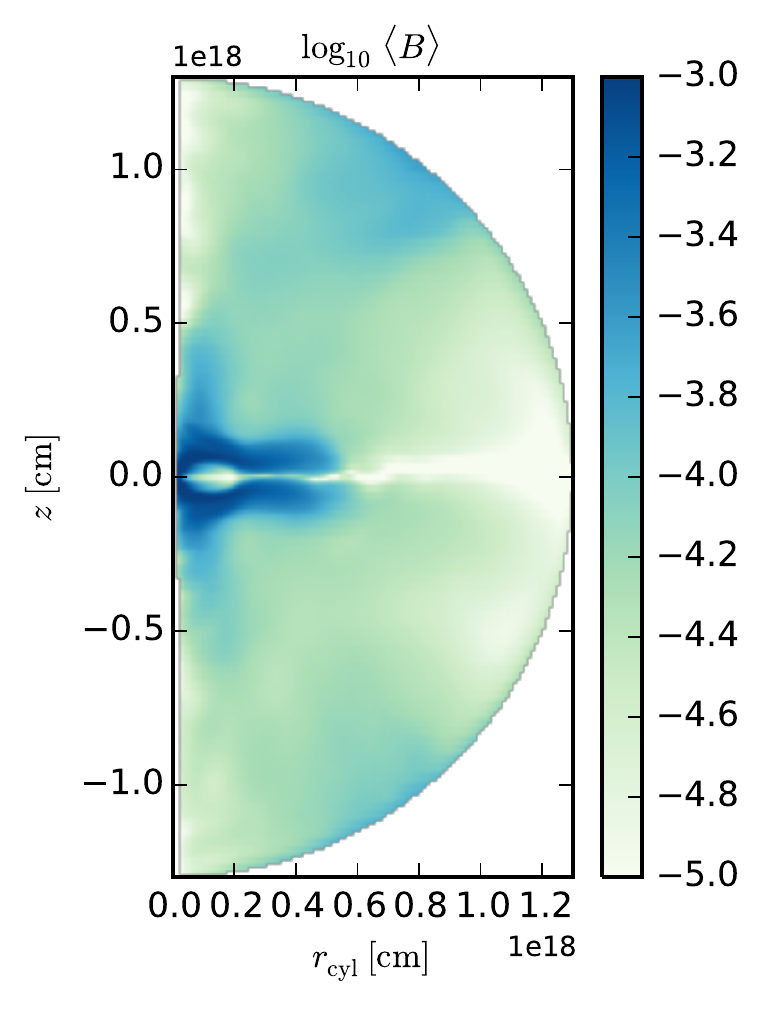}
\includegraphics[height=7.5cm]{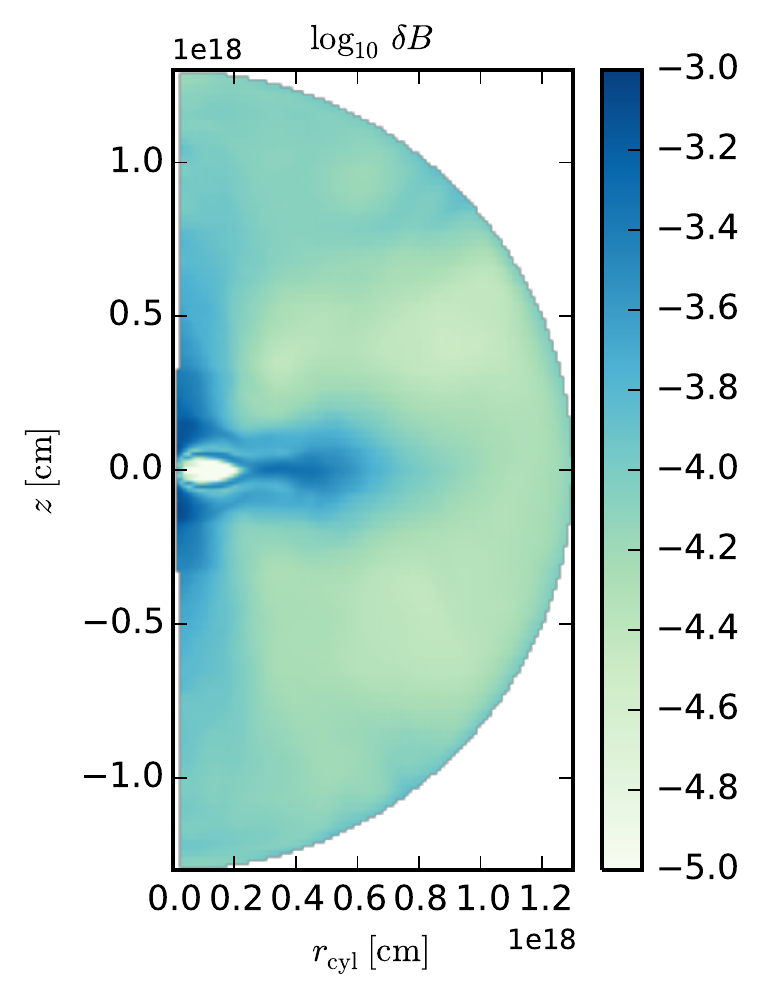}
\includegraphics[height=7.5cm]{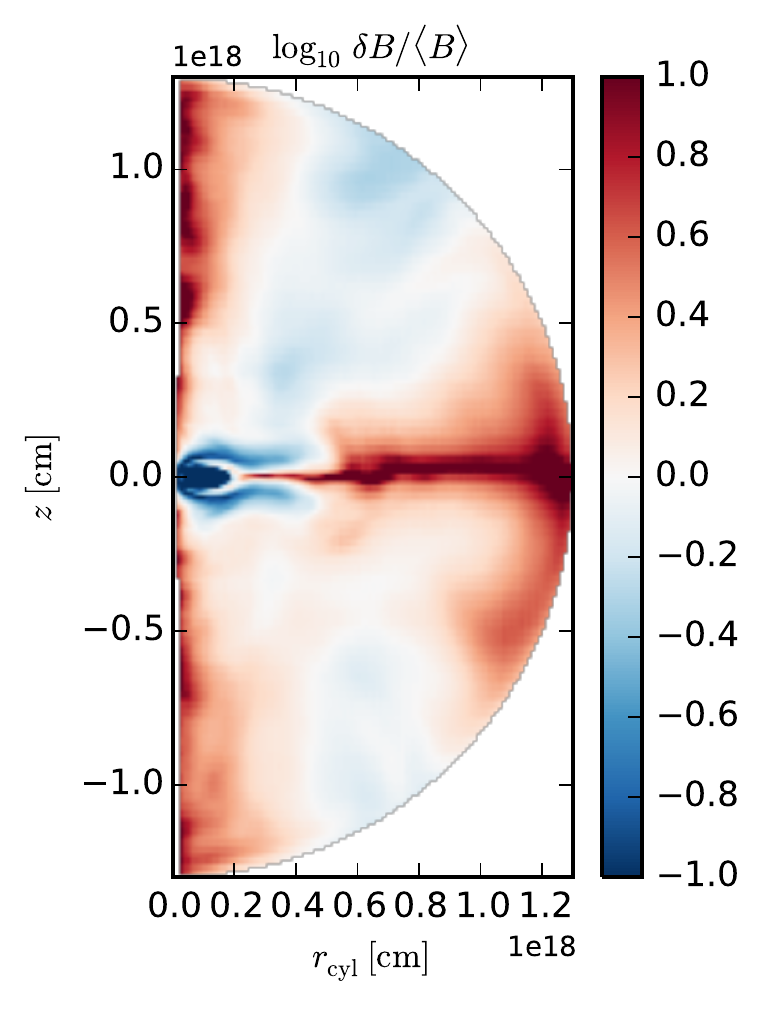}
\includegraphics[height=7.5cm]{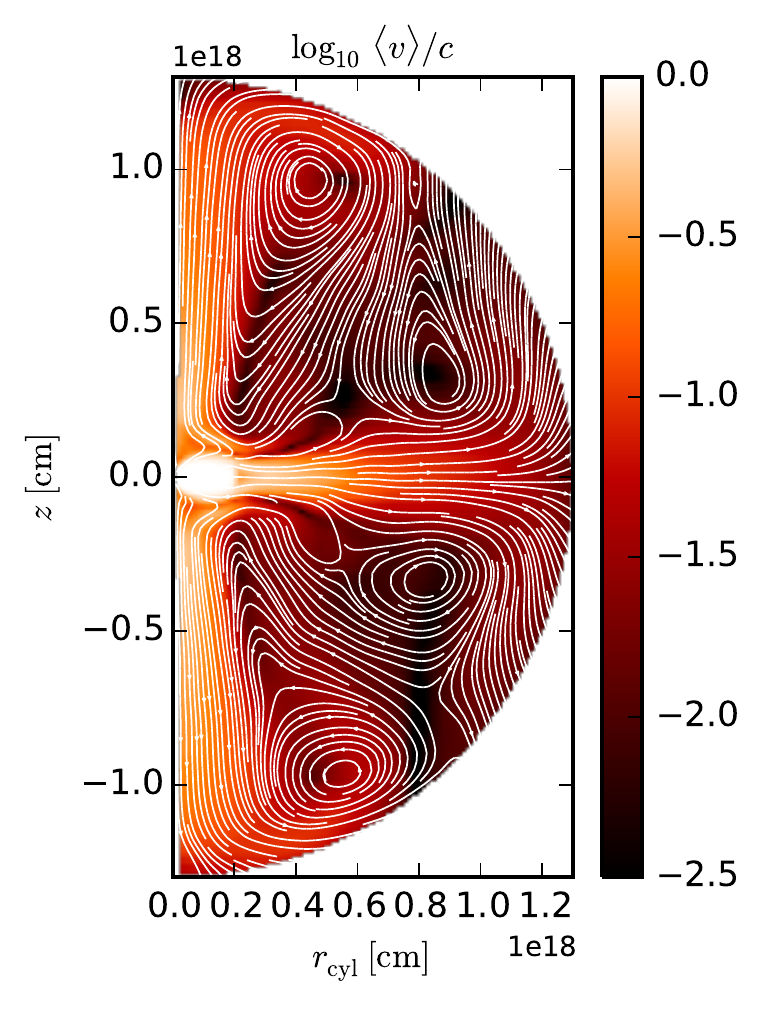}
\includegraphics[height=7.5cm]{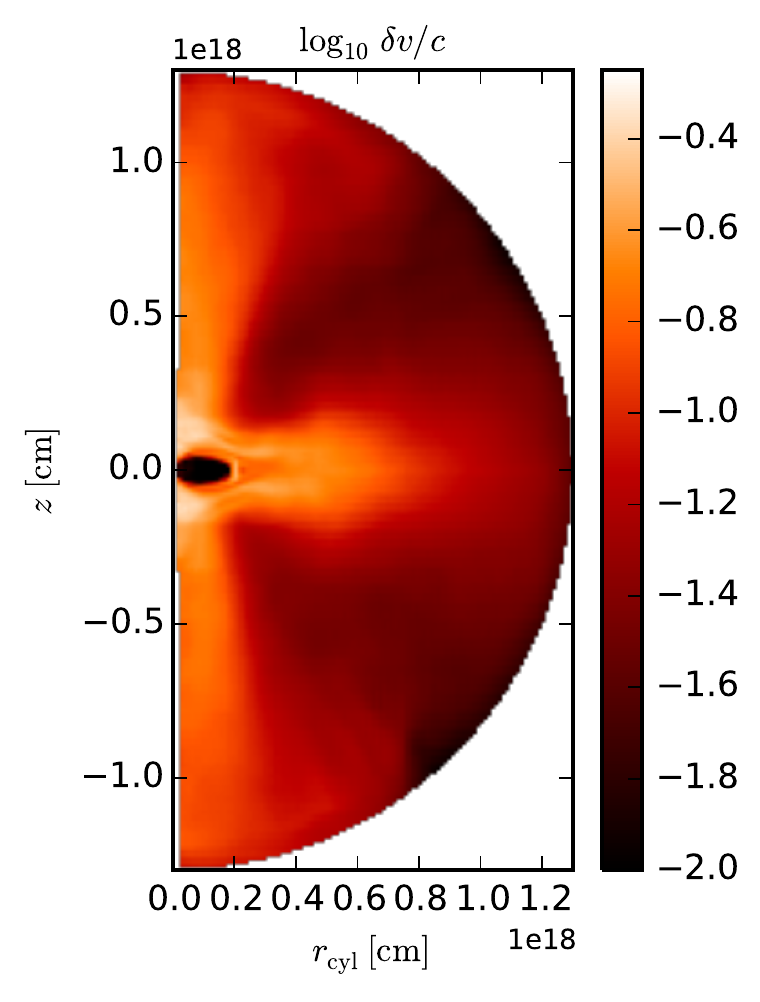}
\includegraphics[height=7.5cm]{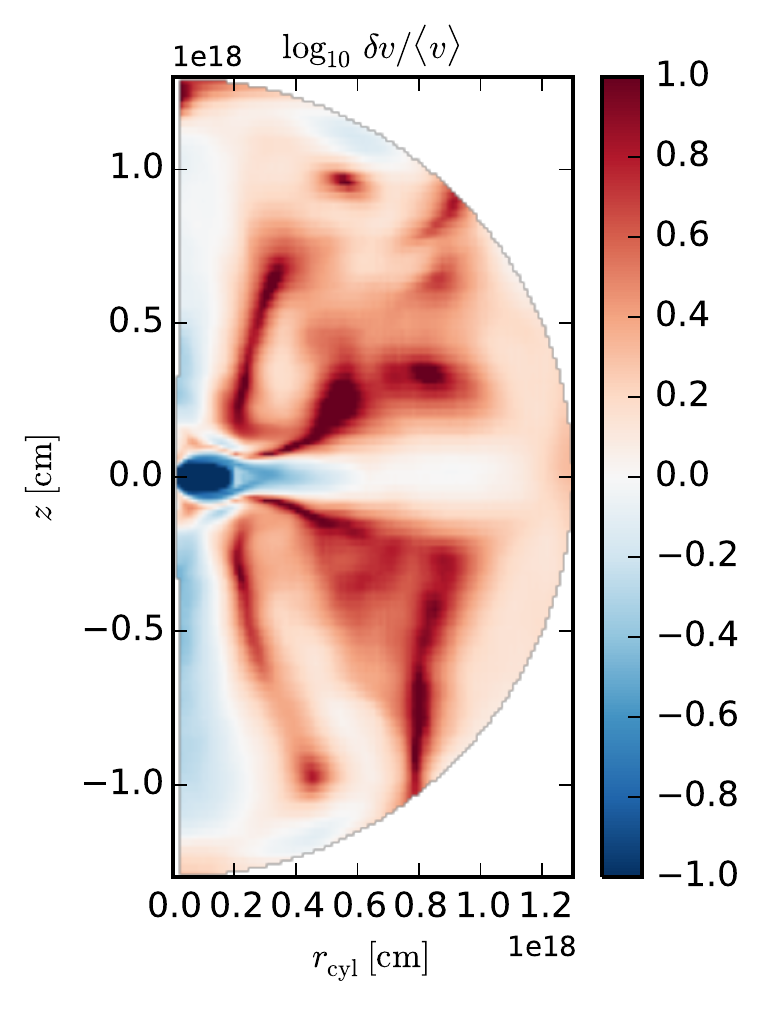}
\caption{Characterisation of the turbulence in magnetic (top) and velocity fields (bottom).
\textit{Left:} Logarithm of the average fields (magnetic field given in Gauss).  We also show stream-lines of the poloidal velocity vector-fields as white lines.  Two large-scale counter-streaming vortices form on each hemisphere.  
\textit{Middle:} Magnitude of the fluctuating field component.    
\textit{Right:} Fluctuating components in terms of the average.  
We average over the azimuthal angle and 21 snapshots, corresponding to 20 years.  }
\label{fig:Dboverb}
\end{center}
\end{figure*}

Maps of the average and fluctuating magnetic field are shown in Fig. \ref{fig:Dboverb}.  As discussed already in \cite{Porth2014}, the magnetic field in the nebula varies by at least a factor of 10 from the area around the termination shock to the outer parts.  The magnetic field strength is lowest in the equatorial and polar regions where changes of polarity occur.  
These are also the regions of the strongest turbulence as characterized by $\delta B/\langle B\rangle$.  The fluctuating component can become more than ten times the average field strength.  
In the more moderate mid-latitude regions of the outer nebula, we still obtain a range of $\delta B/\langle B\rangle\in[0.3-1]$.  
Close to the termination shock however the field is regular and $\delta B/\langle B\rangle\sim 0.1$.  This is true in particular for the flow in the arch-shock where the knot emission is suggested to originate \citep{2015MNRAS.454.2754Y,2016MNRAS.456..286L}.  

Comparing Figs. \ref{fig:Dboverb} with \ref{fig:dvsenergy}, one can see that fluctuations of the magnetic field do not correlate necessarily with regions with strong diffusion of particles as would be suggested by quasi linear theory (QLT) and field line random walk (FLRW) of particle transport \citep[e.g.][]{PhysRevE.85.026411}.  
The high values of $\delta B/B\gtrsim 10$ in the outer equatorial region are not seen as enhancements in the diffusion map.  

The mean velocity in the nebula highlights the fast jet- and equatorial-flow and we recover two large-scale vortices per hemisphere that re-cycle this material.  Comparing the fluctuating component with the average velocity shows that the fluctuations in the re-cycling region are typically stronger than the mean flow and hence the large-scale vortices cannot be a stationary feature.  
In regions of fast systematic flow velocity, fluctuations are relatively weak, however velocity fluctuations are largest with $\delta v/v\gtrsim 10$ in the corresponding shear-layers.  
It is interesting to point out that right at the base of the jet, both velocity and magnetic fluctuations indicate strong turbulence.  
As suggested by e.g. \cite{Porth2014}, this turbulent region should be identified with the highly variable ``anvil'' observed in the Crab nebula \citep{hester1995,hester2002,tavani2011}.  

To test the hypothesis outlined in Sec. \ref{sec:eddydiff} that particles are transported mainly via turbulent flow in the nebula, we apply the Taylor-Green-Kubo (TGK) formulation \citep[e.g.][and references therein]{shalchi2007} directly to the MHD \textit{velocity field}:
\begin{align}
\hat{D}_{xy}(\Delta t) \equiv \int_{0}^{\Delta t} d\tau \langle (v_{x}(\tau)-\langle v_{x} \rangle) (v_{y}(0) - \langle v_{y} \rangle) \rangle \label{eq:tgk}
\end{align}
where we subtract the background flow velocities $\langle v_{x,y} \rangle$. 
In the following, we will only study the radial transport, thus $x,y=r$.  Assuming homogeneity in time and purely diffusive behavior, this definition would be identical to the general form Eq. (\ref{eq:Dxy}) if the velocities were taken as particle velocities.  
On the other hand, as long as the magnetic field remains toroidally dominated, the components of particle drift velocity $\mathbf{(E\times B)~\hat{e}_{r}}$ and flow velocity $v_{r}$ can be interchanged.  It is hence instructive to see how Eq. (\ref{eq:tgk}) compares to the direct measurement from test-particles using Eq. (\ref{eq:Dxy}).  

\begin{figure}
\begin{center}
\includegraphics[height=7cm]{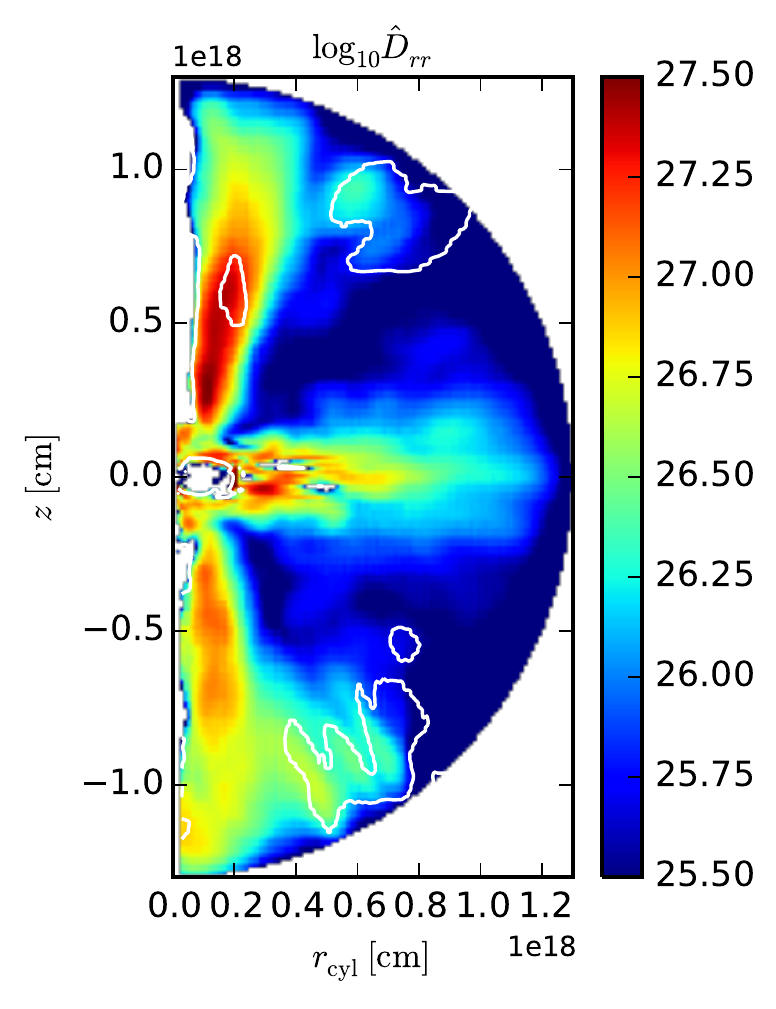}
\caption{
Diffusion resulting from fluctuations of the velocity field according to Eq. (\ref{eq:tgk}) for $\Delta t=5$.  White contours indicate the regions where the radial velocity field is still correlated $(R=1/2)$.  Regions where $\hat{D}_{rr}\le0$ are masked out.  
We average over the azimuthal angle and 21 snapshots, corresponding to 20 years.  }
\label{fig:tgk}
\end{center}
\end{figure}

To obtain convergence of the integral $\hat{D}_{rr}$, we need to ensure that the upper bound $\Delta t$  is chosen larger than the correlation time following from 
\begin{align}
R(\Delta t) = \frac{\langle \left(v_{r}(\Delta t)-\langle v_{r} \rangle\right) \left(v_{r}(0) - \langle v_{r} \rangle\right) \rangle}{\langle v_{r}^{2}\rangle - \langle v_{r} \rangle^{2}} .
\end{align}
The right-hand panel of Fig. \ref{fig:tgk} shows the resulting coefficient for radial diffusion $\hat{D}_{rr}$ after an integration over five years.\footnote{In principle the integration time should be chosen as large as possible.  However, we note that convergence is lost for larger integration times, most likely due to finite box effects.}  We overplot white contours of $R=1/2$ to visually separate correlated from uncorrelated regions.  
At this time, the flow is uncorrelated in most regions which indicates convergence of $\hat{D}_{rr}$.  Qualitatively, we obtain good agreement with the measurement from test particles.  The region of strongest diffusion is near the poles, diffusion is suppressed at polar angles of $45^{\circ}$ and it becomes stronger again in the equatorial region.  
In the polar jet, we also obtain reasonably good quantitative agreement with the test particle simulation and diffusivities reach the maximum of $D_{rr}\approx 3\times 10^{27}\rm cm^{2}s^{-1}$.  

However, there are also clear differences in the results obtained via formulations (\ref{eq:Dxy}) and (\ref{eq:tgk}).  
The total average value $\langle\hat{D}_{rr}\rangle_{r,\theta}\simeq1.2\times 10^{26}\rm cm^{2}s^{-1}$ is almost three times lower than the corresponding test particle result.   
In particular for the outer regions the flow velocity field tends to under-predict $D_{rr}$ by an order of magnitude.  We suggest that this is related to the generation of a poloidal field component such that the radial particle transport decouples from the flow-velocities.  Particles can then be transported outwardly also along the field lines, subject only to pitch-angle scattering.  

\begin{figure}
\begin{center}
\includegraphics[width=0.45\textwidth]{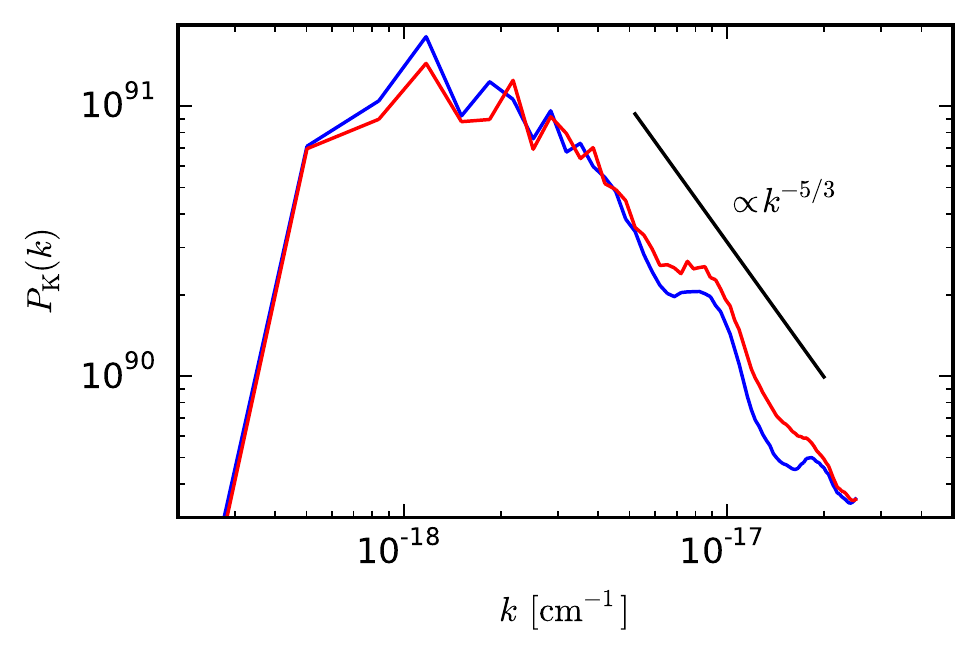}
\caption{
Velocity power-spectrum in the 3D domain for a numerical resolution of $\Delta x=2\times 10^{16}\rm cm$ (blue) and $\Delta x=1\times 10^{16}\rm cm$ (red).  The variable termination shock-radius corresponds to a wave-number of $k_{\rm S}\approx 5\times 10^{-18} \rm cm^{-1}$.  To guide the eye, we also show a Kolmogorov $5/3$ powerlaw which matches the overall cascade quite well.  
}
\label{fig:spec}
\end{center}
\end{figure}

Finally, we investigate the power-spectrum of the flow in the nebula proper.  
Fig. \ref{fig:spec} shows the temporally averaged quantity $P_{\rm K}(k) = \langle \mathbf{\hat{u}\cdot \hat{u}^{*}}\rangle_{t}$ where $\mathbf{\hat{u} = \hat{u}}(k)$ denotes the shell-average of the Fourier-transformed four-velocity.  Note that the we omit the factor $2\pi$ in the wavenumber $k\equiv1/\lambda$ and thus the termination shock corresponds to a wavenumber of $k_{\rm S}\approx 5\times 10^{-18} \rm cm^{-1}$.  
In principle we would expect a peak in the powerspectrum at the driving scale $k_{\rm S}$.  Since we did not subtract the mean velocity, it is difficult to see this in the data as  the spectrum is contaminated by mean-flows on larger scales, e.g. in the jet.  
It is reassuring to see that the large scales are well converged between the two resolutions.  
Both resolutions show a notable bump at $k=10^{-17}\rm cm^{-1}$ which is approximately the transverse size of the fast equatorial flow.  
Apart from this bump, the spectrum on scales smaller than $L_{S}$ can be approximated by a Kolmogorov $5/3$ law (Relativistic turbulence does show  the  classical Kolmogorov scaling \citep{2013ApJ...766L..10R}.) 

\section{The Transport Model}\label{sec:model}

Having derived diffusion coefficients, we now turn our attention to implementing these coefficients in a particle transport model, with the ultimate goal of modelling the X-ray emission observed from three PWNe.  

The electrons and positrons responsible for emitting the observed X-ray emission have a relatively short synchrotron lifetime compared to the age of the nebula, with an estimate for the synchrotron life time is given by \citep[see, e.g.,][]{Dejager2009}
\bb\label{eq:tau_syn}
t_{\rm{syn}} = 1.3 \times 10^{3} \left(\frac{B}{100 \mu \rm G}\right)^{-2} \left(\frac{E}{\rm TeV}\right)^{-1} \rm yr
\ee
For an average magnetic field of $B=300\,\mu G$ and a $1\,\text{TeV}$ electron, the synchrotron lifetime is $t_{\rm{syn}} \sim 100\,\text{yr}$.  This implies that new particles must continually be injected at the termination shock of the PWN.  Consequently one may assume that this particular particle population has reached a steady-state.  Furthermore, as the evolution of the PWN occurs on time scales that are larger than the above-mentioned synchrotron lifetime, one may also assume a steady-state for the magnetic field,  convection velocity, and diffusion coefficient.  Lastly, the X-ray emission will be modelled under the assumption that the PWNe have a spherically symmetric morphology.  The validity of this last assumption varies between PWNe, and will therefore be discussed when the results for the individual sources are presented.    

In a spherically symmetric, steady-state system where convection, diffusion, adiabatic losses, and synchrotron emission are important, the evolution of the lepton spectrum is described by the Fokker-Planck transport equation \citep[see, e.g.][]{Ginzburg1964, Parker1965}
\begin{multline}\label{eq:te}
\kappa\frac{\partial^2 f}{\partial r^2} + \left[\frac{1}{r^2}\frac{\partial}{\partial r}\left(r^2\kappa\right)-V\right]\frac{\partial f}{\partial r} \\
+ \left[\frac{1}{3r^2}\frac{\partial}{\partial r}\left(r^2V\right)+z_p p\right]\frac{\partial f}{\partial \ln p}+4z_ppf  = 0,
\end{multline}
where $r$ is the radial position, $p$ is the particle's momentum, and $f(r,p,t)$ is the omni-directional distribution function.  Furthermore, $\kappa(r)$ is the diffusion coefficient, $V(r)$ the convection velocity, and 
\begin{equation}
z_p(r)=\frac{4}{3}\sigma\s{T}\frac{1}{\left(m_ec\right)^2}\frac{B^2}{8\pi}
\end{equation}
a coefficient related to the synchrotron loss rate.  In the equation above $\sigma\s{T}$ represents the Thomson cross-section, $m_e$ the mass of the electron, and $B(r)$ the magnetic field.  The spatial dependence for $B(r)$, $V(r)$, and $\kappa(r)$ that are used for the modelling will be discussed in more detail in Secs. \ref{sec:KC84} and \ref{sec:PKK14}.  

The electrons and positrons injected at the termination shock are included by employing a suitable boundary condition. The number of particles per momentum interval that flow through the inner radial boundary must be equal to the total number of particles produced per time and momentum interval by the pulsar, $Q_s=Q^*\left(p/p^*\right)^{-(\alpha+2)}$, with $Q^*$ a normalisation constant.  This implies that
\bb\label{eq:bound_flux}
\oint\mathbf{S}_s\cdot d\mathbf{A}_s = Q_s,
\ee
where $d\mathbf{A}_s=r_s^2\sin\theta d\theta d\phi \mathbf{e}_r$ is the surface element on the inner boundary, and 
\begin{equation}
\mathbf{S} = 4\pi p^2\left(Vf-\kappa\cdot\nabla f\right).
\end{equation}
Integrating Eq. (\ref{eq:bound_flux}) leads to the inner boundary condition 
\begin{equation}\label{eq:bound}
CV_s f(r_s,p) - \kappa_s\frac{\partial f(r_s,p)}{\partial r} = Q^*\frac{1}{4\pi r_s^2}\left(\frac{p_s}{p^*}\right)^{-(\alpha+2)}
\end{equation}    
that is solved simultaneously with Eq. (\ref{eq:te}).  

The normalisation constant $Q^*$ is determined by the requirement that the total momentum (energy) in the particle spectrum injected into the nebula must be some fraction $\eta$ of the spin-down luminosity $L$ of the pulsar: 
\bb\label{eq:Q_star}
\int_{p_{\min}}^{p_{\max}} Q^*\left(p/p^*\right)^{-(\alpha+2)} p^3 dp = \frac{\eta L}{c}.
\ee
The value of $\eta$ is left as a free parameter, but it should be noted that this quantity only controls the normalisation of the particle spectrum, and therefore has no influence on the spatial evolution of either the synchrotron photon index $\Gamma$ or the X-ray flux.  Consequently this parameter is only of secondary importance.

Observations indicate that PWNe consist of two particle populations \citep[see, e.g.,][and references therein]{Dejager2009}: the first is responsible for producing the radio synchrotron emission, whereas the second population is responsible for producing the X-ray emission.  The present modelling is concerned with only the latter population, and $p_{\min}$ in Eq. (\ref{eq:Q_star}) therefore represents the minimum particle momentum of the second population.  As the value of $p_{\min}$ is not constrained by observations, $p_{\min} c=0.1\,\text{TeV}$ is used, comparable to the values $p_{\min} c = 0.08-0.15\,\text{TeV}$ derived by \cite{Zhang2008} for three PWNe.  

To calculate the maximum particle momentum $p_{\max}$, or alternatively the maximum energy $E_{\max}$, two possible constraints can be used.  The first is that the rate of acceleration of a particle at a shock will eventually be balanced by synchrotron losses, with \cite{Dejager1996a} deriving the limit
\begin{equation}\label{eq:SynEmax}
E_{\max }  = \gamma _{\max } m_ec^2  = 44 B_s^{-1/2} \text{ erg},
\end{equation} 
where $\gamma_{\max }$ is the maximum Lorentz factor of the accelerated particles and $B_s$ is the magnetic field at the termination shock.  Additionally, particles can only be accelerated as long as they are confined within the shock, leading to the gyroradius limit \citep[see, e.g.,][]{Dejager2009}    
\begin{equation}
E_{\max}=q_e B_s r_s,
\end{equation}
where $r_s$ is the radius of the termination shock.  For the modelling the value of $E_{\max}$ is chosen as the smaller of the two limits.

These limits are based on the assumption that the X-ray emitting electrons and positrons are accelerated through the Fermi mechanism.  One well-known characteristic of this process is that the maximum index of the particle spectrum is $\alpha=2$, with this value chosen for the modelling. 

For the outer boundary it is assumed that the particles can freely escape from the PWN.  A more realistic model would allow for the confinement of particles, but also requires the use of a time-dependent model.  This was investigated by \cite{Vorster2013t}, who found that both boundary conditions lead to identical results.  This makes sense if it is kept in mind that the X-ray emitting particles have a very short lifetime, and one would therefore not expect the number of these particles to increase with time near the outer boundary of the PWN. 

The transport equation (\ref{eq:te}) is solved numerically using the second-order accurate, finite-difference, Crank-Nicolson scheme.  For more details on this model, \cite{Vorster2013a}, and \cite{Vorster2013t} can be consulted, where it is also shown how the various energy loss and transport processes influence the spatial evolution of the particle spectrum.    

As will be discussed in Secs. \ref{sec:KC84} and \ref{sec:PKK14}, the model has a minimum of four free parameters ($B_s$, $V_s$, $K_s$, and $\eta$), and it would thus be too computationally intensive to investigate the whole possible parameter space.  Rather, the search method developed by \cite{Nelder1965} is used to find a set of parameters that result in a minimum $\chi^2$ value.  In order to ensure that the minimum found is not a local minimum, the search method is started at three different positions in the possible parameter space.  

Lastly, for completeness it should be noted that the X-ray data was extracted from either circular or annular regions of increasing size, and that the emission observed from the inner circular regions of the nebula will contain some fraction of the emission emitted from regions further out.  This is taken into account when modelling the data by performing a line-of-sight integral on the emission predicted by the model.

\subsection{The KC84 model}\label{sec:KC84}

This model is base on the well-known results of \cite[][hereafter KC84]{Kennel1984a}.  In this model the magnetic field $B_s$ and the advection velocity $V_s$ at the termination shock are determined by the ratio of electromagnetic to particle energy $\sigma$.  This ratio also determines the spatial dependence of both $B$ and $V$.  However, this model does not provide a direct prediction for the diffusion coefficient.  Furthermore, as a toroidal magnetic field is assumed, diffusion will necessarily occur perpendicular to the magnetic field.  Based on this consideration, the choice of diffusion coefficient is motivated as discussed below.   

It has long been known that the perpendicular diffusion coefficient for charged particles in a collisionless plasma must in some way depend on their Larmor radius, and thus implicitly on the magnitude of the background magnetic field \citep[see, e.g.,][]{Chen1974}. The nature of this magnetic field dependence is, however, not perfectly clear. Many studies assume Bohm diffusion, where $\kappa \propto B^{-1}$ (see Eq. (\ref{eq:bohm})).  This relation has been found, through numerical simulations of perpendicular diffusion coefficients, to be a reasonable choice in the presence of strong turbulence \citep{Hussein2014}. The field line random walk (FLRW) diffusion coefficient represents another limiting case. Here it is assumed that particles exclusively follow magnetic field lines, which themselves wander due to turbulence. The perpendicular diffusion coefficient in this case scales as $B^{-2}$, and is energy-independent \citep[see, e.g.,][and references therein]{Shalchi2009}.  Much theoretical work on perpendicular diffusion has been done in recent years, taking into account more thoroughly the effects of turbulence on perpendicular scattering. These recent theories, being for the most part refinements of the nonlinear guiding center (NLGC) theory proposed by \citet{Matthaeus2003}, tend to yield results that are relatively intractable functions of, amongst other turbulence quantities such as the magnetic variance, the parallel mean free path \citep[see, e.g.,][]{Shalchi2010, Engelbrecht2013, Engelbrecht2015}.  \citet{Shalchi2004}, however, derive approximate, analytical expressions for the NLGC perpendicular mean free path for various limiting scenarios, including the FLRW limit described above, which have been used in heliospheric cosmic-ray modulation studies \citep[see, e.g.,][]{Burger2008}.  These expressions require as input the parallel mean free path. Assuming the quasilinear result for this quantity \citep[see, e.g.,][]{Burger2008}, this perpendicular mean free path will scale as $B^{-4/3}$ at the high energies of interest to this study, and remain energy-independent.   

Based on the above considerations, two scenarios will be investigated.  The first assumes a $B^{-1}$ dependence for the diffusion coefficient (hereafter referred to as the KC84(a) scenario), and the second a $B^{-2}$ dependence (hereafter referred to as the KC84(b) scenario).  These values represent limiting cases, with a more realistic case expected to fall somewhere in between.  Both scenarios additionally assume energy-independent coefficients.  The value of the diffusion coefficient at the termination shock $\kappa_s$ is left as a free parameter, while $\kappa$ is limited by Eq. (\ref{eq:DE}).

\subsection{The PKK14 model}\label{sec:PKK14}  

\begin{figure*}
\begin{center}
  \includegraphics[scale=0.35]{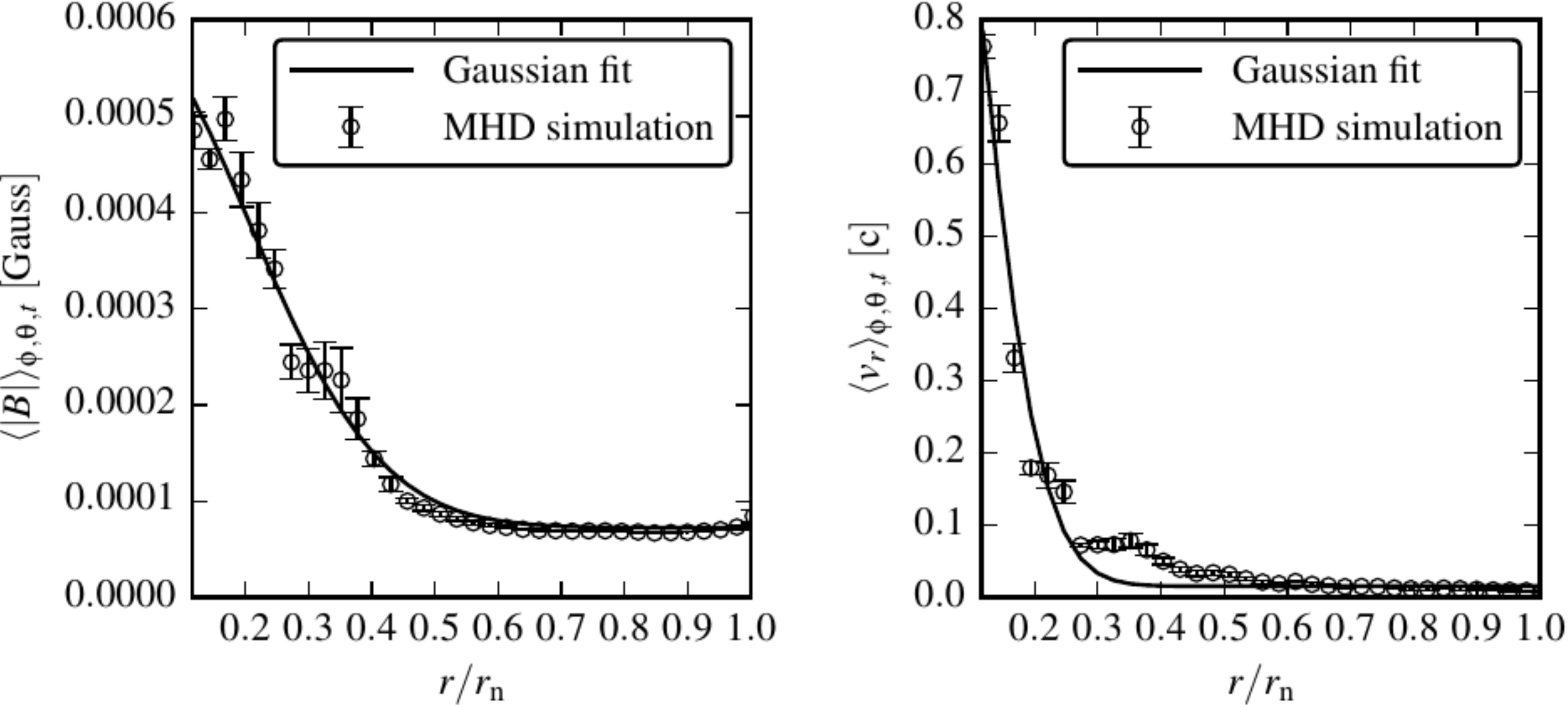}
   \caption{Time-averaged profiles of the magnetic field magnitude and radial velocity using the model ``B3D'' of \citet{Porth2014}.  A Gaussian was fitted to profiles averaged from four snapshots at $t=60, 61, 62, 63$ years after the start of the simulation.}
              \label{fig:V_B_profiles}
\end{center}
\end{figure*}

This model is based on the MHD simulations of \cite[][hereafter referred to as PKK14]{Porth2014}.  To implement these simulations into the spherically symmetric transport model, $B$ and $V$ (predicted by the MHD model) are averaged over time for a number of spherical shells, with the resulting averages shown in Fig. \ref{fig:V_B_profiles}.  The averaged data is fitted with the Gaussian function
\begin{equation}
G(r) = \mathcal{A} \exp\left(-\frac{r^2}{2\mathcal{C}^2}\right)+\mathcal{C},
\end{equation}
with this function having the advantage that it can easily be incorporated into the transport model.  The fits are performed from the termination shock at $r_s \approx 0.15r_{pwn}$ out to the nebula radius $r_{pwn}$ . It yields the parameters $\mathcal{A} = 1.536c$, $\mathcal{B} = 1.56727 \times 10^{−2}c$, and $\mathcal{C} = 0.1005$ for the velocity, and $\mathcal{A} = 5.239 \times 10^{-4}\,\text{G}$, $\mathcal{B} = 7.241 \times 10^{−5}\,\text{G}$, and $\mathcal{C} = 0.2059$ for the magnetic field.  Fits using power-law profiles were also investigated, but it was found that the Gaussian profiles lead to an improvement in the fit.  

As the PWNe that are modelled have varying ratios of $r_s/r_n$, the following characteristics are kept constant for the different PWNe: (1) the value of $B_s$ is chosen as a free parameter; (2) the magnetic field decreases by a factor of $\sim 5$ within approximately three termination shock radii, before reaching a constant value; (3) the velocity at the inner boundary is $V_s=0.51c$; (4) the velocity decreases rapidly within about two termination shock radii, before reaching a constant value; (5) and the constant velocity after two termination shock radii is chosen to match the observed expansion speed of the PWNe.

One aspect of the PKK14 model that has to be investigated is the effect of the rapid decrease in $V$ in the inner regions of the PWN, as this will lead to adiabatic heating.  This requires the use of a time-dependent model that is able to take into account both energy losses and gains.  Using the time-dependent transport model developed by \cite{Vorster2013a} a large number of scenarios with varying parameters were investigated, and it was it was found that adding the effect of adiabatic heating only changes the normalisation of the spectra.  However, the same spatial evolution is predicted, regardless of whether the adiabatic heating in the inner regions of the PWN is included or neglected.  Ideally one would do the fitting with this time-dependent model.  However, compared to the steady-state model it is not only significantly more computationally intensive, but it is also difficult to obtain the same numerical accuracy as the steady-state model.  The fitting will therefore be done using the steady-state model, and neglecting adiabatic heating in the inner regions.

\section{The X-ray Pulsar Wind Nebulae}\label{sec:results}

The three PWNe that are chosen for the modelling are the two young sources G21.5-0.9 and 3C 58, as well as the inner regions of the Vela PWN.  The parameters derived for the nebulae are listed in Tab. \ref{tab:results}.  In order to better compare the results between the models and the selected PWNe, spatially averaged values have also been calculated for the various parameters, and are listed in the second part of the table.  The last line of the table lists the average \emph{P\'{e}clet} number \citep[see, e.g.,][]{Prandtl1953}, defined as
\begin{equation}\label{eq:peclet} 
\xi = \frac{Vr}{\kappa}.
\end{equation}   
This dimensionless number is an indication of the importance of advection relative to diffusion.  When $\xi \gg 1$ the system is advection dominated, while being diffusion dominated for $\xi \ll 1$.      

\begin{table*}
\textbf{\caption[]{\label{tab:results}
{\textnormal{Values derived for the free parameters: \emph{KC84} represent values found using the model of \cite{Kennel1984a}-- Scenario (a) corresponds to the case $\kappa \propto B^{-1}$ and Scenario (b) to the case $\kappa \propto B^{-2}$-- and \emph{PKK14} the values found using the model of \cite{Porth2014}. The first part of the table lists the values of the various parameters at the termination shock, and the second part values averaged over the PWN.}}}}
\begin{center}
	\begin{tabular}{lccccccccc}
	\hline\midrule
	 & \multicolumn{3}{c}{G21.5-0.9} & \multicolumn{3}{c}{Vela} & \multicolumn{3}{c}{3C 58}\\
\cmidrule{2-10}
 	Parameter & KC84(a) & KC84(b) & PKK14 & KC84(a) & KC84(b) & PKK14 & KC84(a) & KC84(b) & PKK14\\
  \midrule
    $B_s$ ($\mu\text{G}$) & $33$ & $33$ & $283$ & $264$ & $264$ & $38$ & $8$ & $8$ & $300$\\
    $V_s$ (units of $c$) & $0.36$ & $0.36$ & $0.51$ & $0.52$ & $0.52$ & $0.51$ & $0.35$ & $0.35$ & $0.51$\\
    $\kappa_s$ ($10^{26}\,\text{cm}^2\,\text{s}^{-1}$) & $8.5$ & $6.0$ & $5.7$ & $0.5$ & $0.5$ & $1.4$ & $27.8$ & $17.1$ & $13.3$\\
    $\sigma$ ($10^{-3}$) & $1.3$ & $1.3$ & $--$ & $142$ & $143$ & $--$ & $0.6$ & $0.6$ & $--$ \\
    $\eta$ ($10^{-2}$) & $8.8$ & $22.5$ & $4.5$ & $0.3$ & $0.3$ & $2.1$ & $--$ & $--$ & $--$\\
  \midrule
    $\bar{B}$ ($\mu\text{G}$) & $158$ & $158$ & $43$ & $30$ & $30$ & $5.8$ & $63$ & $63$ & $46$\\
    $\bar{V}$ ($10^{-3}$, units of $c$) & $4.2$ & $4.2$ & $3.1$ & $159$ & $159$ & $3.3$ & $4.5$ & $4.5$ & $2.6$\\
    $\bar{\kappa}$ ($10^{26}\,\text{cm}^2\,\text{s}^{-1}$) & $1.2$ & $0.3$ & $5.7$ & $1.1$ & $0.9$ & $1.4$ & $1.8$ & $0.3$ & $13.3$\\
    $\bar{\xi}$ & $2.0$ & $9.7$ & $0.3$ & $129$ & $186$ & $2.1$ & $2.0$ & $15$ & $0.2$\\
  \midrule  
\end{tabular}
\end{center}
\end{table*}

\subsection{G21.5-0.9}

With a spin-down luminosity of $L=3.3 \times 10^{37}\,\text{erg}\,\text{s}^{-1}$ \citep{Camilo2006}, the pulsar in the supernova remnant G21.5-0.9 is one of the most energetic pulsars in the Galaxy.  The PWN is located at a distance of $4.8\,\text{kpc}$ \citep{Tian2008}, with an estimated age of $870\,\text{yr}$ \citep{Bietenholz2008}.  X-ray observations \citep{Slane2000, DeRosa2009, Tsujimoto2011} reveal a bright, highly spherical nebula with a radius of $r\s{pwn}\sim 40''$, while \cite{Slane2000} estimated a value of $r_s \gtrsim 1".5$ for the radius of the termination shock.  In order to keep the number of free parameters to a minimum, the value $r_s=1".5$ is used for fitting the data.   

\cite{Bietenholz2008} measured that the PWN is expanding at a velocity of $V\s{pwn}=910\pm 160\,\text{km}\,\text{s}^{-1}$.  With the assumed value of $r_s$, the KC84 model predicts that the above-mentioned expansion velocity in the outer regions of the PWN is compatible with the range of values $1.0\times 10^{-3} \le \sigma \le 1.6\times 10^{-3}$.  For the modelling the intermediate value $\sigma = 1.3\times 10^{-3}$ is chosen, leading to magnetic field strength of $B_s=33\,\mu\text{G}$ at the termination shock.  This range of $\sigma$ values is also comparable to the value $\sigma=3\times 10^{-3}$ derived for the Crab Nebula \citep{Kennel1984b}.  For the PKK14 model, the value $V\s{pwn}=910\,\text{km}\,\text{s}^{-1}$ is used.       

The data that is modelled is the $2-8\,\text{keV}$ observations reported by \cite{Tsujimoto2011}.  For these observations data was extracted from circular regions of increasing size, with the centre of each region placed on the position of the pulsar.  The observations therefore represent cumulative data.  A $3-45\,\text{keV}$ observation for the region $r\le 30''$ has also recently been reported by \cite{Nynka2014}.  The authors found statistically significant evidence for a spectral break at $\sim 9\,\text{keV}$, deriving  $\Gamma=1.852\pm 0.0011$ for the $3-9\,\text{keV}$ energy range, and $\Gamma=2.099^{+0.019}_{-0.017}$ for the $9-45\text{keV}$ energy range.  However, the $3-9\,\text{keV}$ measurement is not entirely in agreement with the results of \cite{Tsujimoto2011}, who found $\Gamma=1.78\pm 0.02$ for the same region.  Despite this discrepancy, the single $9-45\text{keV}$ observation of \cite{Nynka2014} was taken into account for the modelling, as it was found that this does help to constrain the possible free parameters.  As such the single observation should have only a limited influence on the $\chi^2$ value calculated for the fit.       

\begin{figure*}
\begin{center}
  \includegraphics[scale=0.65,angle=-90]{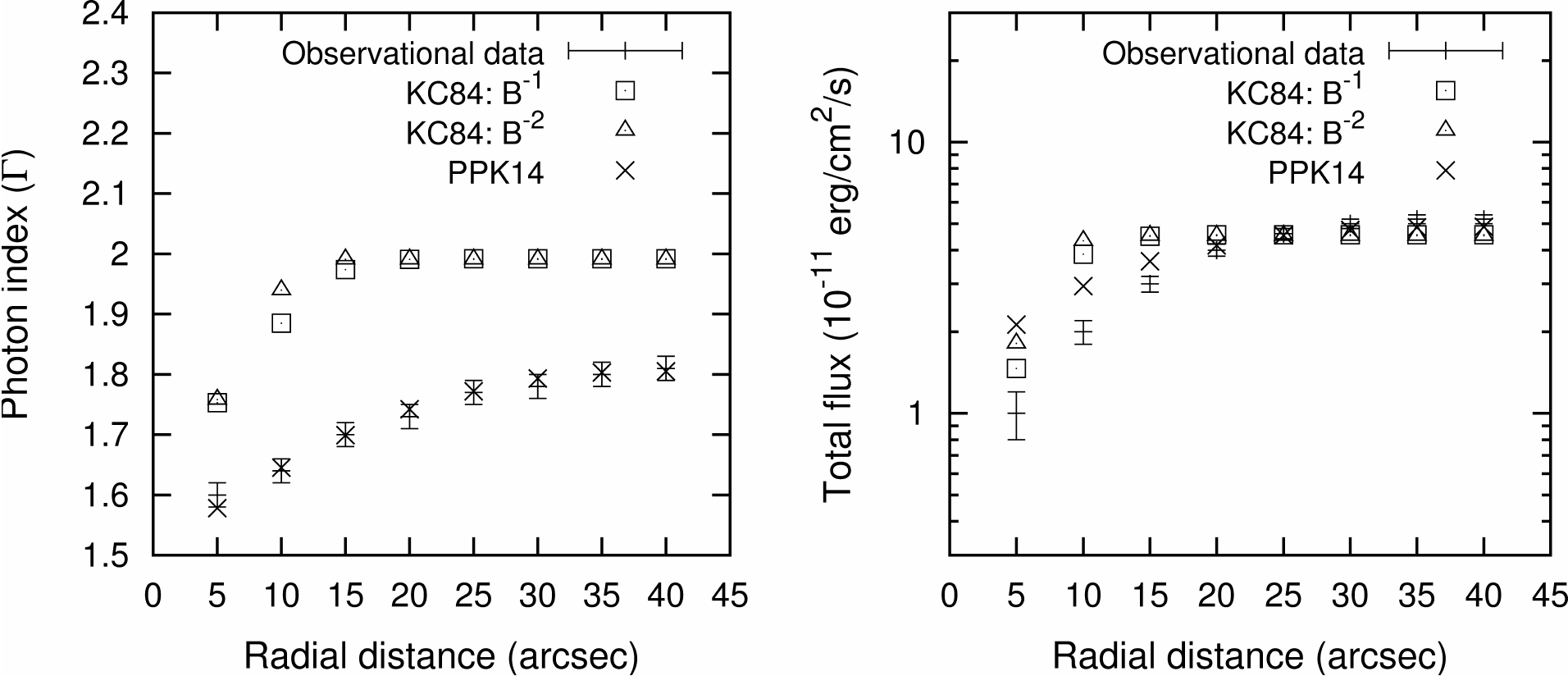}
   \caption{G21.5-0.9: Fit to the data using the magnetic field and velocity profiles from the model of \citet[][PKK14]{Porth2014}, and using the profiles from the model of \citet[][KC84]{Kennel1984a}.  The X-ray data is taken from \citet{Tsujimoto2011} and corresponds to the $2-8\,\text{keV}$ energy range.}
              \label{fig:G21.5-0.9}
\end{center}
\end{figure*}

The best-fit model prediction using the KC84 model is shown in Fig. \ref{fig:G21.5-0.9}, with the best-fit values listed in Tab. \ref{tab:results}.  The figure shows that the KC84 model is not able to reproduce the evolution of $\Gamma$.  Additionally, there is also little variation in the results when a diffusion coefficient is used that scales as either $\kappa\propto B^{-1}$ or $\kappa\propto B^{-2}$ (henceforth referred to as KC84(a) and KC84(b)).  The KC84 model is however capable of modelling the evolution of the flux to some degree, with the largest deviation between the model prediction and the data occurring in the inner parts of the nebula.  The average magnetic field derived from the fit is $\bar{B}=158\,\mu\text{G}$, comparable to the equipartition value $B=180\,\mu\text{G}$ estimated by \cite{Safi-Harb2001}, and used by \cite{Tang2012} in their modelling of G21.5-0.9.  However, using both X-ray and very high energy gamma-ray observations, \cite{Dejager2008d} derived the lower value $\bar{B}=25\,\mu\text{G}$, comparable to the values $\bar{B}=47\,\mu\text{G}$ and $\bar{B}=64\,\mu\text{G}$ derived by \cite{Tanaka2011} using a spatially independent transport model.   

The average diffusion coefficients $\bar{\kappa}=1.2\times 10^{26}\,\text{cm}^2\,\text{s}^{-1}$ and $\bar{\kappa}=2.5\times 10^{25}\,\text{cm}^2\,\text{s}^{-1}$, corresponding to the KC84(a) and KC84(b) scenarios respectively, are smaller than the value $\bar{\kappa}=2\times 10^{27}\,\text{cm}^2\,\text{s}^{-1}$ derived by \cite{Tang2012}.  The average \emph{P\'{e}clet} numbers $\bar{\xi}=2.0$ and $\bar{\xi}=9.7$, for the KC84(a) and KC84(b) scenarios respectively, show that convection is the more dominant particle transport process.  As such one would expect advection to be more dominant for the KC84(b) scenario compared to the KC84(a) scenario: $B$ increases as a function of radius for the largest part of the nebula, and therefore the $B^{-2}$ scaling implies that $\kappa$ should decrease faster than for the KC84(b) scaling.  However, Fig. \ref{fig:G21.5-0.9} shows that the different scalings used for $\kappa$ have only a small influence on the results, as one would expect in a scenario where advection is more dominant.

Fig. \ref{fig:G21.5-0.9} also shows the fit obtained using the PKK14 model, with the best-fit parameters listed in Tab. \ref{tab:results}.  The figure shows that the PKK14 model can reproduce the evolution of $\Gamma$ significantly better than the KC84 model, but has a problem in predicting the correct flux in the inner regions of the nebula.  This is not necessarily surprising as the the flow and magnetic field have a complexity (see Fig. \ref{fig:Drr-rz}) that is not taken into account in the spherically symmetric transport equation.  The average magnetic field derived from the fit is $\bar{B}=43\,\mu\text{G}$, comparable to the value derived from the KC84 model.  However, it may be argued that the value $B_s=283\,\mu\text{G}$ predicted by the PPK14 model is more realistic given the large luminosity of the pulsar powering the G21.5-0.9.  The average diffusion coefficient $\bar{\kappa}=5.7\times 10^{26}\,\text{cm}^2\,\text{s}^{-1}$ is once again limited by Eq. (\ref{eq:DE}).  As is the case for the KC84 model, the PKK14 model predicts that diffusion should be the dominant transport process, with the value $\bar{\xi}=0.34$ calculated from the fit.  Lastly, the PKK14 model predicts that $\Gamma=1.89$ in the $9-45\text{keV}$ energy range.

\begin{figure*}
\begin{center}
  \includegraphics[scale=0.65,angle=-90]{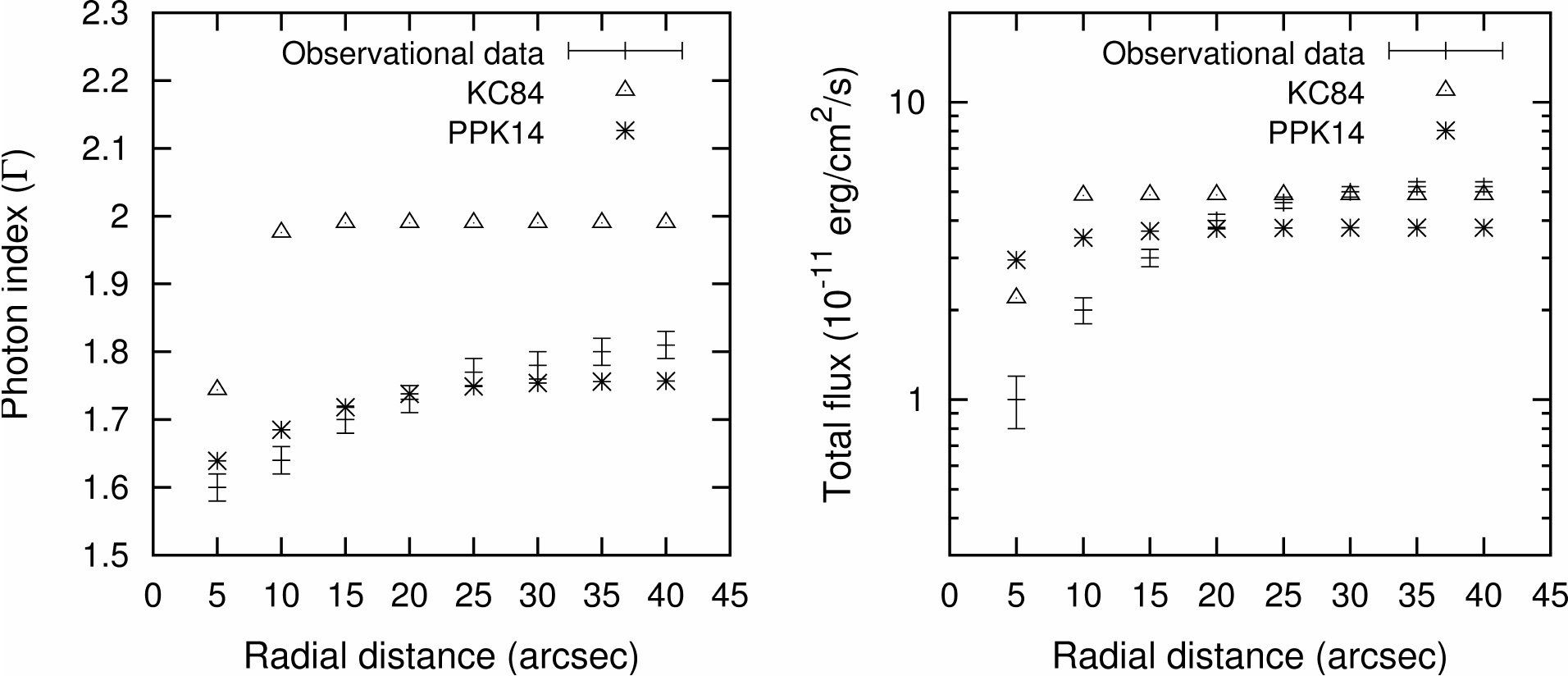}
   \caption{The same as Fig. \ref{fig:G21.5-0.9}, but with diffusion neglected.}
              \label{fig:G21.5-0.9_no_diffusion}
\end{center}
\end{figure*}

To illustrate the role of diffusion, it is useful to fit the data with diffusion neglected.  The results of this comparison is shown in Fig. \ref{fig:G21.5-0.9_no_diffusion}.  As such it is difficult to make any conclusions based on the results from the KC84 model.  Not only does the model fail to fit the evolution of $\Gamma$, but the values $\xi > 1$ in Tab. \ref{tab:results} predict that convection should be more important that diffusion.     

Within the context of the PKK14 model, comparison of Figs. \ref{fig:G21.5-0.9} and \ref{fig:G21.5-0.9_no_diffusion} shows that diffusion leads to a moderately better fit to $\Gamma$, but a significantly better fit to the evolution of the flux.  However, there are two important aspects to keep in mind.  The first is that, in order to improve the fit to the data, the spectral index injected at the shock was changed to $\alpha=2.2$, i.e., neglecting diffusion can to some degree be compensated for by changing certain parameters.  Secondly, and more importantly, the average magnetic field in the absence of diffusion is $\bar{B}=6.9\,\mu\text{G}$, lower than the prediction of any previous model.  The consequence is that the conversion efficiency of the pulsar's spin-down luminosity to particle energy must exceed $100\%$, i.e., $\eta > 1$. 

The discussion presented in the previous paragraph also implicitly illustrates an important consequence of diffusion.  In order for electrons and positrons to reach the outer regions of G21.5-0.9, the transport time scale has to be shorter than the synchrotron life time of the particles.  Neglecting diffusion increases the transport time scale, and subsequently the synchrotron life time of the particles has to decrease, implying that $\bar{B}$ has to become smaller.  Thus, with diffusion included it becomes possible for the particles to propagate to distances from the pulsar that would otherwise not have been possible due to synchrotron losses.

\subsection{The inner regions of the Vela PWN}

The Vela PWN, located at a distance of $290\,\text{pc}$ \citep{Dodson2003b}, is a highly asymmetric PWN, as shown in X-ray \citep{Markwardt1995} and very high energy gamma-ray observations \citep{H_Aharonian2006a}, with the age of the nebula estimated to be $11400\,\text{yr}$ \citep{Manzali2007}.  Although the Vela PWN is a morphologically complex source with an extension of $\sim1^{\circ}$, $3-10\,\text{keV}$ observations \citep{Mangano2005} of the inner region show a bright X-ray nebula that is approximately spherically symmetric.  X-ray observations also show a double torus and jet-like structure at the centre of the PWN \citep{Helfand2001}, commonly associated with the radius of the termination shock.  Using a geometrical model to fit the double torus, \cite{Ng2004} derived a value of $r_s\sim 21"$.  The aim is not to model the entire PWN, but only the above-mentioned inner region.  The $3-10\,\text{keV}$ data \citep{Mangano2005} that is modelled was again extracted from circular regions of increasing size up to a radius of $r=15'$.

As the velocity $V$ is unknown, the value of $\sigma$, and consequently $B_s$, is now an additional free parameter in the KC84 model.  For the PKK14 model, it is assumed the velocity reaches the constant value $V=1000\,\text{km}\,\text{s}^{-1}$ after two termination shock radii (see also discussion in Sec. \ref{sec:PKK14}), comparable to the value measured for G21.5-0.9.          

\begin{figure*}
\begin{center}
  \includegraphics[scale=0.65,angle=-90]{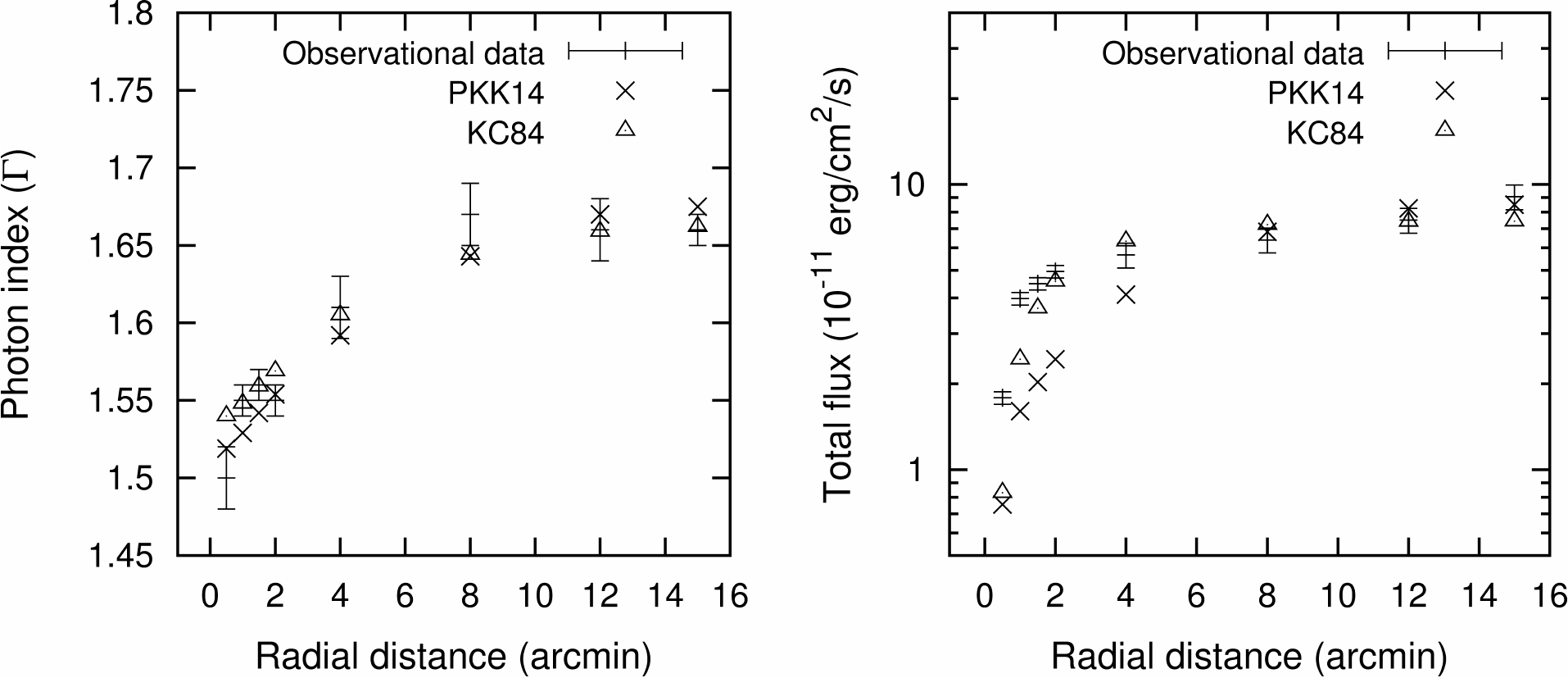}
   \caption{Inner regions of Vela: Fit to the data using the PKK14 and KC84 models.  The X-ray data is taken from \citet{Mangano2005} and corresponds to the $3-10\,\text{keV}$ energy range.}
              \label{fig:Vela}
\end{center}
\end{figure*}

Fig. \ref{fig:Vela} shows the fits to the data using the KC84 model.  It can be seen that the model can fit the $\Gamma$ data moderately well, while providing a good fit to the flux data for the largest part of the nebula.  However, there is a discrepancy between the model prediction and flux data in the outer regions of the nebula.  Both $\kappa$ scenarios ($\kappa \propto B^{-1}$ and $\kappa \propto B^{-2}$) predict an average magnetic field of $\bar{B}=30\,\mu\text{G}$ for the inner region of the nebula, larger than the average magnetic field of $\bar{B}=5\,\mu\text{G}$ estimated by \citet{Dejager2008b} for the whole PWN.  The best-fit value $\sigma=0.14\times 10^{-1}$ is also in agreement with the range $\sigma=0.05-0.5$ estimated by \citet{Sefako2003}, and comparable to the range $\sigma=0.01-0.1$ estimated by \citet{Bogovalov2005}.  Average diffusion coefficients of $\bar{\kappa}=1.1\times 10^{26}\,\text{cm}^2\,\text{s}^{-1}$ and $\bar{\kappa}=8.7\times 10^{25}\,\text{cm}^2\,\text{s}^{-1}$ are predicted for the $\kappa\propto B^{-1}$ and $\kappa\propto B^{-2}$ scenarios respectively.  These values correspond to P\'{e}clet numbers of $\bar{\xi}=130$ and $\bar{\xi}=186$, indicating that advection is the dominant process. 

A salient feature of Fig. \ref{fig:Vela} is the similarity between the two $\kappa$ scenarios in the KC84 model.  The large value of $\sigma$ implies that $B$ decreases in the inner regions of the PWN, before approaching a constant value.  The constant value therefore implies that the radial dependence of $\kappa$ will also be constant, regardless of the scaling used.     

Apart from the $3-10\,\text{keV}$ data, \cite{Mangano2005} also extracted a single data point from a $r=15'$ circular region in the $20-100\,\text{keV}$ energy range.  The photon index and flux was measured as $\Gamma=2.00\pm 0.05$ and $F_{X}=15.9\pm 0.4\,\text{erg}\,\text{cm}^{-2}\,\text{s}^{-1}$, respectively.  This data was also taken into account for the modelling, with the KC84 model fit in Fig. \ref{fig:Vela} predicting $\Gamma=1.78$ and a flux of $F_{X}=10.6\,\text{erg}\,\text{cm}^{-2}\,\text{s}^{-1}$ for both diffusion coefficients.

Fig. \ref{fig:Vela} also shows that while the PKK14 model can produce a better fit to the $\Gamma$ data, compared to the KC84 model, the former model has trouble in predicting the correct flux in the inner regions of the nebula. A smaller average magnetic field of $\bar{B}=5.8\,\mu\text{G}$ and P\'{e}clet number $\bar{\xi}=2.1$ is found for the PKK14 model, although a diffusion coefficient $\bar{\kappa}=1.4\times 10^{26}\,\text{cm}^2\,\text{s}^{-1}$ comparable to those of the KC84 model is predicted.  Lastly, the PKK14 model predicts $\Gamma=2.01$ and a flux of $F_{X}=17.2\,\text{erg}\,\text{cm}^{-2}\,\text{s}^{-1}$ in the $20-100\,\text{keV}$ energy range.  Note that this is an improvement over the prediction made by the KC84 model (cf. previous paragraph).

\subsection{3C 58}

The PWN 3C 58, located at a distance of $3.2\,\text{kpc}$ \citep{Roberts1993}, is generally associated with the supernova SN 1181, implying an age of $833\,\text{yr}$ \citep[see, e.g.,][]{Slane2004}.  X-ray observations in the $2.2 - 8\,\text{keV}$ energy range not only reveal a nebula with an elliptical morphology similar to that of the Crab Nebula, but also a show jet-like and a possible torus structures in the inner regions of the PWN \citep{Slane2004}.  The torus structure is elongated, and extends between $2''.5$ and $8''.0$ from the pulsar \citep{Slane2004}, indicating that the torus is inclined with respect to the plane of the sky.  To derive an estimation of the termination shock radius requires the use of a geometrical model similar to the one presented by \cite{Ng2004}.  As this is beyond the scope of the present modelling, the average value $r_s=5''.25$ is used.

Radio measurements show that the nebula is expanding at a velocity of $V\s{pwn}\sim 910\pm 360\,\text{km}\,\text{s}^{-1}$ and $V\s{pwn}\sim 550\pm 220\,\text{km}\,\text{s}^{-1}$ along the major and minor axes of the nebula, respectively.  The X-ray measurements were extracted from circular annuli centred on the pulsar, and the average expansion velocity $V\s{pwn}=730\,\text{km}\,\text{s}^{-1}$ is thus used, implying $\sigma=5.55\times 10^{-4}$ and $B_s=8.4\,\mu\text{G}$ for the KC84 model.  Unfortunately normalised flux measurement are not available, and the fitting was therefore done based on the spatial variation of only $\Gamma$.

\begin{figure}
\begin{center}
  \includegraphics[scale=0.65,angle=-90]{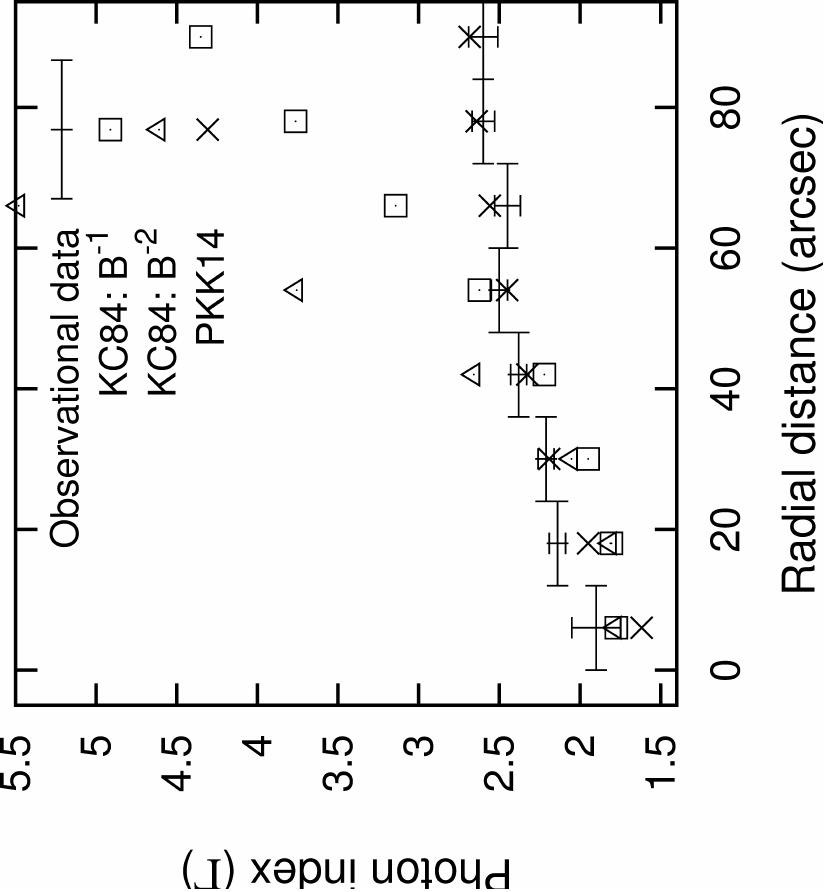}
   \caption{3C 58: : Fit to the data using the PKK14 and KC84 models.  The X-ray data is taken from \citet{Slane2004} and corresponds to the $2.2-8\,\text{keV}$ energy range.}
              \label{fig:3C_58}
\end{center}
\end{figure}

Fig. \ref{fig:3C_58} shows the KC84 model can reproduce the $\Gamma$ observations in the very inner regions of the PWN, but fails to reproduce the data for the largest part of the nebula.  By contrast, the PKK14 model is able to predict the evolution of $\Gamma$.  The value of $\bar{B}=46\,\mu\text{G}$ predicted by the PKK14 model is comparable to a previous estimate $\bar{B}=40\,\mu\text{G}$ derived by \cite{Tanaka2013}, but larger than the value of $\bar{B}=14\,\mu\text{G}$ estimated by the MAGIC collaboration following their detection of 3C 58 at TeV gamma-ray energies \citep{Aleksic2014}.  Note that the above-mentioned values are all smaller than the estimated equipartition value $\bar{B}=80\,\mu\text{G}$ \citep{Green1992}.

\section{Discussions and Conclusions}\label{sec:conclusions}

Turbulence in PWNe gives rise to high diffusive particle transport.  This idea has been previously employed by \cite{Tang2012}, who found that the synchrotron spectral index of several young PWNe could be modelled well by relying on diffusive transport alone (i.e., neglecting advection).  In the model of \cite{Tang2012} the diffusion coefficient $\kappa$ was left as a free parameter, whereas the present paper aims to constrain $\kappa$ using three-dimensional MHD simulations of PWNe \citep{Porth2014}.  The MHD results are then used as input in a particle transport model, with the ultimate goal of modelling the X-ray emission from G21.5-0.9, the inner regions of the Vela PWN, and 3C 58.   

Comparing the results from MHD simulations that contain an integrated test-particle component with the diffusion coefficient derived from a formalism based on correlations of the flow velocity field, we conclude that MHD turbulence gives rise to high diffusivities that are important for particle transport.  
For typical parameters of young PWNe, we obtain radial diffusion coefficients in excess of $10^{26}\rm cm^{2}s^{-1}$.  
Motivated by these results, we suggest a simple scaling relation between the termination shock size $L_{\rm s}$ and the diffusion coefficient in the PWN:  
$D_{rr}\propto v_{\rm f}L_{\rm s}$,
where $v_{\rm f}\approx0.5 c$ is a typical flow velocity at the driving scale corresponding to $L_{\rm s}$ (see Eq. (\ref{eq:DE})).  
Both in our simulated PWN and in real systems, the driving is localised in the centre of the nebula, with the turbulence decaying as one moves away from the centre.  This would lead to a lower radial transport in the outer regions of the nebula, which is however compensated for by the formation of a radial component in the magnetic field.  

The transport is split into advection with the mean flow plus a diffusive contribution due to advection in the turbulent flow component.  
While both contributions are ``advective'', only the mean flow is associated with adiabatic losses in the modeling.  
Not surprisingly, we do not obtain a significant energy dependence in the diffusion coefficient for particles with gyroradii smaller than $L_{s}$.  
Thus for average field strengths of $B_{0}<300 \mu\rm G$ and shock radii $L_{s}> 0.1 \rm ly$, we can safely neglect any energy dependence even for particles emitting MeV synchrotron radiation.  The outward particle transport is clearly non-Bohm.

Diffusive transport is most efficient in the kink unstable polar jet where $D_{rr}> 10^{27}\rm cm^{2} s^{-1}$, marking good agreement between the test-particle approach and results based on correlations of the flow field alone.  
By contrast, the corresponding Bohm diffusion coefficient for X-ray emitting particles is orders of magnitude lower, on the order of $5\times 10^{23} \rm cm^2 s^{-1}$  in a $100\mu \rm G$ field.  
In the outer equatorial and mid-latitude regions we find stronger discrepancies between the two approaches where the flow field underestimates the diffusive transport.  
We suggest that the development of a poloidal component to the magnetic field in these outer regions is responsible for the increased efficiency of diffusive transport.  

To model the observed spatial evolution of the X-ray photon spectra of the three PWNe G21.5-0.9, the inner regions of Vela, and 3C 58, a steady-state, spherically symmetric transport model that includes convection, diffusion, adiabatic cooling, and synchrotron losses is used.  This model has as input time averaged radial profiles for the magnetic field, velocity, and diffusion coefficient that are obtained from the MHD model.  In order to better understand the results, the modelling was also done using the magnetic field and velocity profiles from the well-known MHD model of \cite{Kennel1984a}.  As diffusion is not a feature of the latter model, two scenarios were considered.  In the first $\kappa \propto B^{-1}$, and in the second $\kappa \propto B^{-2}$.  The model of \cite{Kennel1984a} is also useful to derive $\sigma$, the ratio of magnetic to particle energy in the nebula (in the MHD model of \citet{Porth2014}, $\sigma = 1$) 

It was found that using the magnetic and velocity profiles of \cite{Porth2014} allows one to find reasonable fits to the evolution of the photon index $\Gamma$, whereas using the profiles from the \citet{Kennel1984a} model leads to fits that are, in general, not as good.  However, the model of \citet{Kennel1984a} does lead to somewhat better fits to the the evolution of the X-ray synchrotron flux.  However, taking into account the fits to both the spectral index and flux, the MHD model of \citet{Porth2014} is the preferred model.  

Using the \citet{Porth2014} model, it is found that diffusion is the more important transport mechanism in G21.5-0.9 and 3C 58, whereas advection is the more important mechanism in the inner regions of the Vela PWN.  However, for all three sources the P\'eclet number is close to unity, indicating that both advection and diffusion play an important role in particle transport.

One noteworthy point is that while the model of \citet{Kennel1984a} has trouble in fitting the spectral indices of G21.5-0.9 and 3C 58, the model does allow one to find a reasonable fit to the photon index and flux data of the inner regions of the Vela PWN.  From these fits the value $\sigma = 0.14$ is derived, significantly higher than the value of $\sigma=3\times 10^{-3}$ derived for the Crab Nebula \citep{Kennel1984b}, with this latter value often taken as a canonical value for PWNe.  However, the larger $\sigma$ value for Vela is not entirely unexpected as \citet{Sefako2003}, and \citet{Bogovalov2005} have estimated comparable values.

For further studies, we suggest performing simulations with a larger dynamic range between the driving scale and the domain size in order to obtain better statistics on the turbulent flow.  It is also important to check the results against advanced cosmic ray turbulence theories developed for the solar system \citep[e.g.][]{2008JGRA..113.8105B,Zank2012} as the presence of subsonic motion in the PWN and large relativistic velocity fluctuations could render them quite different.  
Apart from spatial diffusion, future studies will also investigate diffusion in momentum space with special attention to Fermi-II type acceleration in the nebula proper.

\section*{Acknowledgements}
This work was partially funded through 
the STFC under the standard grant ST/I001816/1, 
the ERC Synergy Grant ``BlackHoleCam'' (Grant 610058) 
and the NSF grant number 1306672.

\bibliographystyle{aa}
\bibliography{References_Vorster_Lyutikov_2014,astro}

\end{document}